\documentclass[12pt]{article}

\usepackage{amsmath,amssymb,graphicx,multirow,xspace,makeidx}
\usepackage[colorlinks=true,urlcolor=blue,anchorcolor=blue,citecolor=blue,filecolor=blue,linkcolor=blue,menucolor=blue,pagecolor=blue]{hyperref}
\usepackage[compress,numbers]{natbib}
\usepackage{subfigure, placeins}
\usepackage{booktabs}
\usepackage{blindtext, rotating}
\usepackage{afterpage}
\usepackage{enumitem}
\usepackage{ marvosym }
\usepackage{verbatim}
\usepackage{authblk}
\usepackage{xcolor}

\makeindex

\newcommand{\newc}{\newcommand}
\newc{\gsim}{\lower.7ex\hbox{$\;\stackrel{\textstyle>}{\sim}\;$}}
\newc{\lsim}{\lower.7ex\hbox{$\;\stackrel{\textstyle<}{\sim}\;$}}
\newc{\gev}{\,{\rm GeV}}
\newc{\mev}{\,{\rm MeV}}
\newc{\ev}{\,{\rm eV}}
\newc{\kev}{\,{\rm keV}}
\newc{\tev}{\,{\rm TeV}}

\def\ln{\mathop{\rm ln}}

\def\Im{\mathop{\rm Im}}
\def\Re{\mathop{\rm Re}}

\newc{\mz}{M_Z}
\newc{\mpl}{M_*}
\newc{\mw}{m_{\rm weak}}
\newc{\nr}[1]{N^c_R{}_{#1}}
\usepackage{amsmath}

\def\bitem{\begin{itemize}}
\def\eitem{\end{itemize}}
\newc{\ie}{{\it i.e.}}          \newc{\etal}{{\it et al.}}
\newc{\eg}{{\it e.g.}}          \newc{\etc}{{\it etc.}}
\newc{\cf}{{\it c.f.}}

\newcommand{\CO}{\mathcal{O}}

\newcommand\fverb{\setbox\fverbbox=\hbox\bgroup\verb}
\newcommand\fverbdo{\egroup\medskip\noindent%
            \fbox{\unhbox\fverbbox}\ }
\newcommand\fverbit{\egroup\item[\fbox{\unhbox\fverbbox}]}
\newbox\fverbbox

\numberwithin{equation}{section}

\long\def\symbolfootnote[#1]#2{\begingroup%
\def\thefootnote{\fnsymbol{footnote}}\footnote[#1]{#2}\endgroup}

\newcommand{\be}{\begin{equation}}
\newcommand{\ee}{\end{equation}}

\newcommand{\bea}{\begin{eqnarray}\begin{aligned}}
\newcommand{\eea}{\end{aligned}\end{eqnarray}}
\newcommand{\mat}{\begin{pmatrix}}
\newcommand{\rix}{\end{pmatrix}}

\renewcommand{\bar}{\overline}

\newcommand{\qq}{\qquad}

\newcommand{\beqa}{\begin{eqnarray}}
\newcommand{\eeqa}{\end{eqnarray}}
\newcommand{\lp}{\left(}
\newcommand{\rp}{\right)}

\newcommand{\beq}{\begin{equation}}
\newcommand{\eeq}{\end{equation}}

\newcommand{\abs}[1]{\left\vert#1\right\vert}

\newcommand{\order}[1]{{\cal O}\left(#1\right)}

\newcommand{\CA}{\mathcal{A}}

\def\FF{$\mathtt{FormFlavor}$}
\def\FFM{$\mathtt{FFModel}$}
\def\FFW{$\mathtt{FFWilson}$}
\def\FFO{\texttt{FFObservables}}
\def\FFP{\texttt{FFPackage}}
\def\CA{$\mathtt{CalcAmps}$}
\def\mma{$\mathtt{Mathematica}$}
\def\FA{$\mathtt{FeynArts}$}
\def\FC{$\mathtt{FormCalc}$}
\def\SF{$\mathtt{SUSY\_Flavor}$}
\def\FK{$\mathtt{FlavorKit}$}
\def\SP{$\mathtt{SPheno}$}
\def\code#1{\ensuremath{\texttt{#1}}}

\def\file#1{\ensuremath{\mathtt{#1}}}

\newcommand{\amd}{``Acc'' mode}
\newcommand{\fmd}{``Fast'' mode}

\newcommand{\block}[1]{\begin{center} \vspace{-6mm}  \hspace{5mm} \fcolorbox{black!10}{black!10}{\begin{minipage}[c]{0.95\textwidth}
#1\end{minipage} 
}\end{center} }
\newcommand{\blockc}[1]{\block{\begin{center} #1\end{center} }}
\newcommand{\codebox}[3]{\block{\begin{tabular}{rl}
\code{#1} & \begin{minipage}[c]{#3\textwidth}#2\end{minipage}
\end{tabular}}}

\begin{document}

\title{\FF\ Manual
}

\date{\today}

\author[1]{Jared A.~Evans}
\author[2]{David Shih}

\affil[1]{\small{Department  of Physics\\ 
University of Illinois at Urbana-Champaign\\ 
Urbana, IL 61801}}
\affil[2]{NHETC\\ 
Department of Physics and Astronomy\\
Rutgers University\\
Piscataway, NJ 08854 }

\maketitle

\begin{abstract}
This manual describes the usage and structure of \FF, a \mma-based tool for computing a broad list of flavor and CP observables in general new physics models. Based on the powerful machinery of \FA\ and \FC, \FF\ calculates the one-loop Wilson coefficients of the dimension 5 and 6 Standard Model effective Lagrangian entirely from scratch. These Wilson coefficients are then evolved down to the low scale using one-loop QCD RGEs, where they are transformed into flavor and CP observables. The last step is accomplished using a model-independent, largely stand-alone  package called \FFO\ that is included with \FF.  The SM predictions  in \FFO\ include up-to-date references and accurate current predictions. Using the functions and modular structure provided by \FF, it is straightforward to add new observables. Currently, \FF\ is set up to perform these calculations for the general, non-MFV MSSM, but in principle it can be generalized to arbitrary \FA\ models.  \FF\ and an up-to-date manual can be downloaded from: \url{http://formflavor.hepforge.org}.

\end{abstract}

\newpage

\tableofcontents

\newpage

\section{Introduction}
\label{sec:intro}

Precision flavor and CP observables, such as $\Delta m_K$, $\epsilon_K$ and BR$(B\to X_s\gamma)$, have long been invaluable probes of physics beyond the Standard Model (BSM) (see e.g.\ the review \cite{Buras:1998raa}). While the Standard Model enjoys approximate flavor and CP symmetries that keep these observables small, new physics generally does not. Indeed, many models of new physics predict deviations at some level from Standard Model expectations.  A notable example is supersymmetry (SUSY), which is highly motivated on many grounds, yet generically predicts large contributions to flavor- and CP-violating observables (see e.g.\ \cite{Martin:1997ns} for a review and original references).  

With new, high-precision experimental results in flavor physics on the horizon,  such as those promised by Belle II \cite{Abe:2010gxa}, LHCb \cite{Alves:2008zz}, and NA62 \cite{Fantechi:2014hqa,NA62TDR}; and lattice calculations evolving into an era of higher and higher precision \cite{Aoki:2013ldr}, there is increasing need for equally precise theoretical tools to facilitate the exploration of constraints on new physics models. In general, theoretical predictions for flavor observables are derived from the Wilson coefficients of dimension 5 and dimension 6 effective operators built out of Standard Model fields. There are many such operators, and, in many models (e.g., SUSY), these Wilson coefficients do not arise at tree-level, necessitating the calculation of one- or even higher-loop diagrams. With the many loop functions and Wilson operators involved, performing an accurate assessment of the flavor constraints on even a single parameter point in the MSSM across a dozen distinct observables is an onerous task to perform by hand.  Although several publicly available programs exist to calculate flavor and CP observables, these often assume minimal flavor violation (MFV), lack a sufficiently broad list of flavor observables, are numerically unstable, or contain bugs, likely introduced in transcribing loop formulas by hand from the literature.  

It was with these issues in mind that \FF\ was designed.  \FF\ is a \mma-based, general-purpose tool for computing a broad list of flavor and CP observables in new physics models. \FF\ is built on the powerful machinery of \FA\ \cite{Hahn:2000kx} and \FC\ \cite{Hahn:1998yk}, which facilitate the automatic generation and evaluation of Feynman diagrams for general Lagrangians. Using \FA\ and \FC, \FF\ calculates the one-loop Wilson coefficients from scratch, greatly improving the reliability of the code. Currently, \FF\ is set up to perform these calculations for the general, non-MFV MSSM, but in principle it can be generalized to arbitrary models. \FF\ contains two distinct running modes, Fast and Accurate, allowing for a safer evaluation of Passarino-Veltmann integrals without suffering from numerical instabilities.

\begin{figure}[!t]
\begin{center}
\begin{picture}(395,215)(0,0)
\put(100,130){\fcolorbox{black!20}{black!02}{\begin{minipage}[c]{0.57\textwidth}
{\bf FFPackage \vspace{56mm} } \end{minipage}}}
\put(6,135){\fcolorbox{black!10}{black!10}{\begin{minipage}[c]{0.120\textwidth}
{\bf FeynArts}
\end{minipage}}}
\put(5,96){\fcolorbox{black!10}{black!10}{\begin{minipage}[c]{0.125\textwidth}
{\bf FormCalc}
\end{minipage}}}
\LARGE
\put(-20,51){\fcolorbox{black!10}{black!10}{\begin{minipage}[c]{0.235\textwidth}
{\bf CalcAmps}
\end{minipage}}}
\normalsize
\put(20,10){\fcolorbox{black!10}{black!10}{\begin{minipage}[c]{0.310\textwidth}
{\color{red}analytic Wilson coefficients}
\end{minipage}}}
\normalsize
\put(110,142){\fcolorbox{black!10}{black!10}{\begin{minipage}[c]{0.160\textwidth}
{CompileAmps}
\end{minipage}}}
\LARGE
\put(225,137){\fcolorbox{black!10}{black!10}{\begin{minipage}[c]{0.220\textwidth}
{\bf FFWilson}
\end{minipage}}}
\LARGE
\put(120,184){\fcolorbox{black!10}{black!10}{\begin{minipage}[c]{0.420\textwidth}
{\bf FFModel (MSSM)}
\end{minipage}}}
\normalsize
\put(205,99){\fcolorbox{black!10}{black!10}{\begin{minipage}[c]{0.310\textwidth}
{\color{red}numeric Wilson coefficients}
\end{minipage}}}
\LARGE
\put(201,57){\fcolorbox{black!10}{black!10}{\begin{minipage}[c]{0.330\textwidth}
{\bf FFObservables}
\end{minipage}}}
\normalsize
\put(235,19){\fcolorbox{black!10}{black!10}{\begin{minipage}[c]{0.195\textwidth}
{\color{red}flavor constraints}
\end{minipage}}}
\put(237,230){\fcolorbox{black!10}{black!10}{\begin{minipage}[c]{0.173\textwidth}
{\color{blue}input spectrum}
\end{minipage}}}
\Huge
\put(30,112){$\downarrow$}
\put(29.5,74){$\downarrow$}
\put(28,26){$\Downarrow$}
\put(146,28){$|$}
\put(146,52){$|$}
\put(146,76){$|$}
\put(146,100){$|$}
\put(143.5,120){$\uparrow$}
\put(186,139){$-$}
\put(201.5,139){$\rightarrow$}
\put(143.5,158){$\downarrow$}
\put(273,158){$\downarrow$}
\put(272,114){$\Downarrow$}
\put(273.5,77){$\downarrow$}
\put(272,34){$\Downarrow$}
\put(273.5,208){$\downarrow$}
\end{picture}
\end{center}
\caption{Schematic illustration of the \FF\ code.  The left branch is \CA, which, in principle, only needs to be run once per model; the middle is the compiling portion of \FF, which must be run once per session; to the right is the core code of \FF, which is run once per parameter point evaluated.  In blue is the primary input, a spectrum.  In red are the primary outputs of \FF: the analytic Wilson coefficients, numerical Wilson coefficients, and flavor constraints.  The four main pieces of the code: \FFO, \FFW, \FFM\ (MSSM), and \CA,  are discussed in sections \ref{sec:obs}, \ref{sec:wilson}, \ref{sec:MSSM}, and \ref{sec:CalcAmpsP}, respectively.   \CA\  requires \FA\ and \FC, but the rest of the code does not.
}
\label{fig:FlowChart}
\end{figure}
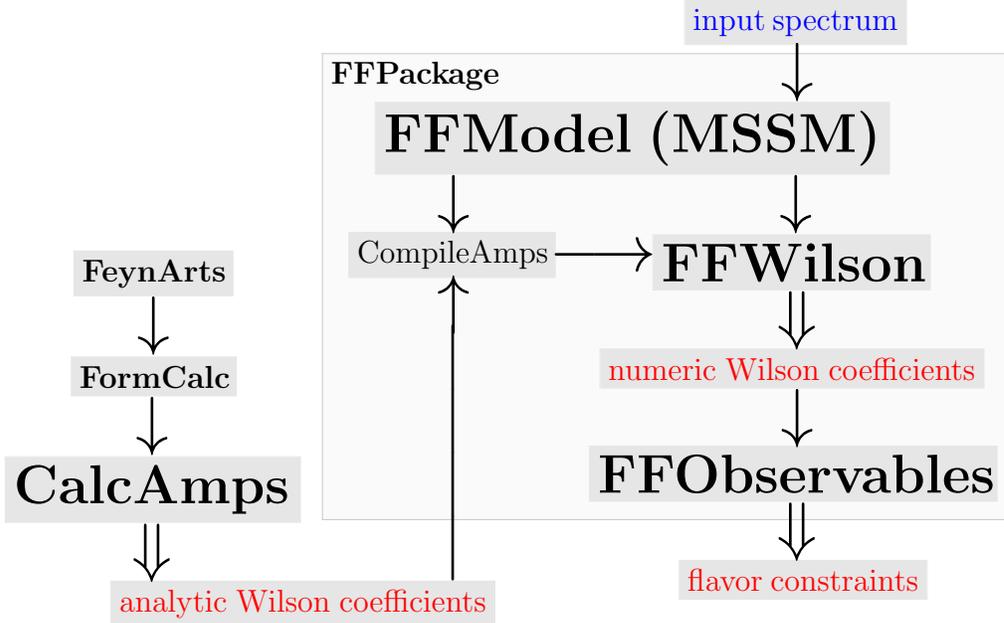

The structure of the \FF\ code is illustrated by the flowchart in fig.~\ref{fig:FlowChart}.
\FF\ can be viewed as two distinct programs, \CA\ and \FFP, and \FFP\ contains a number of separate modules with different functionality.

\begin{itemize}

\item \CA\ automatically generates one-loop amplitudes from a \FA/\FC\ model file and extracts analytic expressions for the Wilson coefficients.  \CA\ only needs to be run once per model in principle, or re-run whenever the user wishes to add an observable. (For the default observables and the default model -- the general flavor- and CP-violating MSSM -- \CA\ need not be run at all; rather the user can use the amplitude files that come with \FF.)

\item \FFP\ contains the  core code of \FF, which is run repeatedly per \mma\ session in order to turn input spectra into flavor and CP observables. The modules of \FFP\ include:

\begin{itemize}
\item \FFW\ takes input spectra and numericizes the analytic Wilson coefficients generated by \CA. It also compiles the analytic Wilson coefficients for faster numerical evaluation. The compilation step is taken care of by \code{CompileAmps}\ and needs to be run only once per \mma\ session.

\item \FFO\ converts numerical Wilson coefficients into flavor and CP observables.    \FFO\ is entirely model independent, and could in principle be run as a standalone package, given a list of Wilson coefficients and the scale where they are defined. Observables are treated in a modular way in \FFO, making it straightforward to add new ones to \FF. This is described in section \ref{sec:newobs}. 

\item \FFM\ contains all of the model-specific code used by \FF. This includes code to read in input spectra in a user-specified format, and code to link the \CA\ output with \FFW. Although the only \FFM\ file that currently exists is the fully general, non-MFV MSSM, in principle, other \FA\ models can be used to generate one-loop amplitudes.  A user only needs to write new \FFM\ modules, at which point the existing \FFW\ and \FFO\ machinery can evaluate them. 

\end{itemize}
\end{itemize}

\FF\ does not treat any higher-loop contributions, such as the double Higgs penguins \cite{Buras:2001mb} or Barr-Zee diagrams \cite{Barr:1990vd} that can be important for $\Delta m_{B_s}$ and the neutron EDM, respectively. Unlike the publicly available program, \SF\ \cite{Crivellin:2012jv}, \FF\ does not account for chirally-enhanced contributions,  important mostly at large $\tan\beta$,  that enter at 2-loops or higher. 

In this manual, we will first present a simple QuickStart guide in section \ref{sec:BUG}.  For the user who wants to acquire flavor constraints from an SLHA2 file, this QuickStart guide contains all of the information needed.  Section~\ref{sec:ffpackage} introduces the main \FF\ package. Section~\ref{sec:obs} presents the details of \FFO, and how all of the observables contained in \FF\ are evaluated.  All details of \FFW\ are discussed in section~\ref{sec:wilson}.   The details of the single \FFM\ included with the public release,  the non-MFV MSSM model, are presented in section~\ref{sec:MSSM}.  The \CA\ package is described in section~\ref{sec:CalcAmpsP}, along with a tutorial on adding new observables to \FF.   A comparison of \FF\ with the public codes \SF\ \cite{Crivellin:2012jv} and \FK\ \cite{Porod:2014xia} is presented in the appendix.

\FF\ has been tested with \mma\ 8, 9, and 10 in Mac OS X 10.9.5 (Mavericks) and Mac OS X 10.10.5 (Yosemite).  All comments on speed throughout the text concern \mma\ 9 with Mac OS X 10.10.5 using 8 GB of RAM and a 2.9 GHz Intel Core i5.

\section{Basic User's Guide}
\label{sec:BUG}

 The basic use of \FF\ takes as input an SLHA2 file in the flavor- and CP-violating MSSM and computes the one-loop contribution to various flavor observables.   Contributions from diagrams containing only standard model particles utilize hardcoded, detailed treatments extracted from the literature, while the one-loop MSSM contributions are evaluated from scratch. In order to combine the two,   the new physics Wilson coefficients must be RG evolved from the SUSY scale down to the relevant low scale for the flavor observable of interest.
 
In this section, we present a brief QuickStart guide that will illustrate how to take the package out of the box and evaluate flavor observables from an SLHA2 file.   For many users, it is likely that the information presented in this section is all of the functionality of \FF\ that will be needed.

\subsection{Starting the Program}

\FF\ can be downloaded from: \url{http://formflavor.hepforge.org}.  Once downloaded, the tar ball should be unpacked.  No installation of the package is necessary.  

  The \FF\ program is loaded using a front-end \mma\ notebook, such as the \file{FormFlavor.nb} notebook provided with the package.    Before loading, the path of \FF\ must be specified  with:
  \blockc{\code{FormFlavor`\$FFPath=\{PATH\}}}  \index{\$FFPath}\index{FormFlavor`\$FFPath}
   The package can be loaded using the \mma\ command (\texttt{Get}):
  \blockc{\code{<<FormFlavor`\$FFPath<>"FormFlavor`"}}
   Within a few seconds, the package should load, listing \FF's version number, the \FFM's version number, and several different portions of the code that were loaded.    No other \mma\  packages (such as \FA\ or \FC) need to be loaded in order for the code to work. 
    
\subsection{Compiling the Amplitudes}
\label{sec:compile}

 Once loaded, many of the code's routines are already functional.  Notably, all of the \FFO\ capabilities can be run (see section \ref{sec:obs} for details).  However, the main operation of \FFW, calculating numerical Wilson coefficients from MSSM parameters, needs to be compiled before it will function.  This compiling takes time and has two separate running modes (``Fast'' and accurate, ``Acc'').  Although the differences between these two modes are discussed in detail in section \ref{sec:LIL}, the basic difference is that \fmd\ evaluates results more quickly, but is subject to occasional numerical instabilities in Passarino-Veltman integral calculations due to the use of double precision, while \amd\ (accurate mode) is substantially slower, but is more resistant to these instabilities. \fmd\ is good for performing parameter scans; \amd\ is good for evaluating individual points. 
 
  When \FF\ is loaded, all existing observable amplitude files are stored in \code{\$FFAmpFileList}\index{\$FFAmpFileList}.   Running:
   \blockc{\code{CompileFF[\$FFAmpFileList]} or \code{CompileFF[\$FFAmpFileList,"Fast"]}}    
   will compile the code in \fmd.  Trading ``Fast'' for ``Acc''  will compile in ``Acc'' mode. 
  
  \codebox{CompileFF[amplist]}{compiles all amplitudes in \code{amplist}  in \fmd}{0.66} \vspace{-6.9mm}
  \codebox{CompileFF[amplist,mode]}{ as above, but for  running mode \code{mode}\index{CompileFF}}{0.55}
   
 Compiling the amplitudes takes several minutes.  Although only one mode needs to be compiled for the program to function, both running modes can be simultaneously loaded.  After compiling a particular mode, the default running mode is set to that mode.  This can be changed at any time by setting \code{\$FFActiveRunningMode="Fast" or "Acc"}\index{\$FFActiveRunningMode}.   However, we note that compiling these large amplitudes stores them into memory, which can consume several GB of RAM.   Compiling both modes simultaneously effectively doubles this RAM consumption.  By decreasing the number of processes loaded, especially the larger memory hogs, such as $B_q\to\mu\mu$, the memory expenditure can be reduced.
 
 Optionally, one can pre-evaluate many of the time-consuming operations performed in the compiling step by building the amplitudes (section \ref{sec:building}).  This exchanges disk space to cut down compiling time by more than factor of two.

\subsection{Calculating Observables}
\label{sec:FFeval}

Once the code has been compiled, the flavor observables can be calculated from the SUSY-scale  parameters of an SLHA2 file by using the command:
 \codebox{ FFfromSLHA2[file]}{read in \code{file} at the SUSY scale and output flavor observables using current active running mode}{0.71}
 \vspace{-4.1mm}
 \codebox{FFfromSLHA2[file,mode]}{ as above, but for  running mode \code{mode}\index{FFfromSLHA2}}{0.60}
  This function returns a nested list of flavor observables in \mma\ list format as,
  \block{\center{\{\{observable name, observable value\},...\}}}

 \begin{table}[t!]
 \begin{center}
 \begin{tabular}{ccc}
status & exclusion value & explanation \\
\hline 
0 & $\frac{X_{FF}-X_{exp}}{\sqrt{ \sigma_{exp}^2+ \sigma_{th}^2}}$ & existing theoretical prediction and measurement \\
1 & $\frac{X_{FF}}{X_{exp,UB}}$ & an experimental upper bound only  \vspace{1.4mm} \\
2 & $\frac{X_{FF}}{X_{exp}+2 \sigma_{exp}}$& experimental measurement, but no reliable theory prediction
\end{tabular}
\end{center}
\caption{The definitions of the exclusion values outputted by \code{FFConstraintsfromSLHA2} \&  \code{FFConstraints} (which is defined in section \ref{sec:ffpackage}), according to the observable status.  Here, $X_{FF}$ is the standard model + new physics value as determined by FormFlavor for the observable $X$, $X_{exp}$ ($X_{exp,UB}$) is the current experimental measurement's central value (upper bound), $\sigma_{exp}$ is the experimental uncertainty on the measurement of $X$, and $\sigma_{th}$ is the theoretical uncertainty on the standard model prediction for $X$.   The treatment of these observables is defined in the function \code{FFObsMeasure}\index{FFObsMeasure} within \file{Core/FFPackage.m}.
\label{tab:constraints}
}
\end{table}

  In order to get the flavor constraints, one can use the command:
  \codebox{FFConstraintsfromSLHA2[file]}{read in \code{file} at the SUSY scale and give flavor constraints with current active running mode}{0.56} \vspace{-4.1mm}
  \codebox{FFConstraintsfromSLHA2[file,mode]}{as above, but for running mode \code{mode}\index{FFConstraintsfromSLHA2}}{0.5}
  The constraints from this are presented in a list of the form
 \blockc{\{\{observable name, exclusion value, status\},...\}}
 where the definition of exclusion value depends on the current experimental/theoretical status of the observable. These definitions are summarized in table~\ref{tab:constraints}. A status of ``0'' corresponds to the common situation where there is both a reliable theoretical prediction in the standard model and an experimental measurement -- example: BR$(B\to X_s \gamma)$.  For observables with this status, an exclusion value with a magnitude greater than two represents a roughly 95\% CL exclusion.   A status of ``1'' is a situation where there is an experimental upper bound only -- example:  the neutron EDM.  For many of these observables, no standard model prediction is included in the \FF\ calculation as these values are much lower than the current experiment measurement.   For observables of this class, an exclusion value with a magnitude larger than one represents a spectrum excluded at roughly 90\% CL.  A status of ``2'' corresponds to a situation where there is an experimental measurement, but no existing, reliable theoretical prediction -- example: $\Delta m_D$. For observables with this status, an exclusion value much larger than one represents a substantial tuning between the unknown (or poorly predicted) standard model contributions and the new physics contributions.

\section{The Main Package: FFPackage}
\label{sec:ffpackage}

In this section we will describe the contents of \FFP, which is the main wrapper package of \FF, 
responsible for loading all other packages (except \CA, see section \ref{sec:CalcAmpsP}) and defining the main \FF\ routine that turns input spectra into flavor and CP observables.

\FF\  has many parameters common to multiple observables, for example, the CKM matrix.  In order to simplify the process of updating these parameters with improved measurements, most parameters are stored in a single file,  \file{Core/SMParameters.m}, that is automatically loaded by \FFP.  Within this file there are conversion factors from SI units to natural units,  standard model particle masses at different scales and often in different schemes, electroweak parameters such as the Higgs vacuum expectation value, Fermi's constant $G_F$, and $\sin\theta_W$, meson masses, decay constants, lifetimes, and branching ratios, and lepton lifetimes.  All parameters are given in natural units of GeV$^n$. Note that \FF\ will ignore SM parameters provided in an SLHA2 file, choosing to use  the hardcoded values provided in \file{Core/SMParameters.m}.
\codebox{GetSMParameter["name"]}{returns SM parameter \code{"name"} \\ type \code{??GetSMParameter} for a complete list\index{GetSMParameter}}{0.65} 

\FFP\ also establishes the basis of Wilson operators used by \FF:
\beqa
\label{eq:OpA}
\CO_A^M(f_1,f_2) &=& e \bar f_1 \sigma^{\mu\nu}P_M f_2 F_{\mu\nu} \\
\label{eq:OpG}
 \CO_G^M(f_1,f_2) &=& g \bar f_1 \sigma^{\mu\nu}P_M f_2 G_{\mu\nu} \\
 \label{eq:OpS}
\CO^{MN}_S(f_1,f_2,f_3,f_4) &=& (\bar f_1 P_M f_2) (\bar f_3 P_N f_4) \\
\label{eq:OpV}
\CO^{MN}_V(f_1,f_2,f_3,f_4) &=& (\bar f_1 \gamma^\mu P_M f_2) (\bar f_3 \gamma_\mu P_N f_4)\\
\label{eq:OpT}
\CO^{MN}_T(f_1,f_2,f_3,f_4) &=& (\bar f_1 \sigma^{\mu\nu} P_M f_2) (\bar f_3 \sigma_{\mu\nu} P_N f_4)
\label{eq:opend}
\index{Operators}
 \eeqa
  where $M,N=L,R$ with $P_{R}=\frac12\lp 1+\gamma_5\rp$ and $P_{L}=\frac12\lp 1-\gamma_5\rp$.    All colors are contracted within the bilinear when such a Fierz rearrangement is possible.  Although currently no operators are used in \FF\ that cannot be rearranged, this information is encoded by the \FC\ notation of:
 \codebox{SUNT[Col1,Col2] SUNT[Col3,Col4]}{ color contractions within the fermion bilinears, i.e., $(\bar f_{1,\alpha} X_M f_{2,\alpha}) (\bar f_{3,\beta}X_N f_{4,\beta})$ \index{SUNT}}{0.5} \vspace{-1mm}
 \codebox{SUNT[Col1,Col4] SUNT[Col3,Col2]}{ color contractions outside of the fermion bilinears, i.e., $(\bar f_{1,\alpha} X_M f_{2,\beta}) (\bar f_{3,\beta}X_N f_{4,\alpha})$}{0.5}
\noindent  All Wilson operators in \FF\ are expressed as   
\blockc{\code{OpX[$P_1$,$P_2$][$\{f_1,f_2\},\{f_3,f_4\}$]} or  \code{OpY[$P_1$][$\{f_1,f_2\},\{v\}$]}} 
for $X=S,\,V,\,T$ or $Y=A,\,G$, $P_i=``L",``R"$, $v=``\gamma",\, ``g"$ $f_i=``b",\,``d",$ etc.  As examples \code{OpV[$``L"$,$``L"$][$\{``s",``d"\},\{``s",``d"\}$]} is the $\CO_V^{LL}$ for kaon mixing and \code{OpA[$``R"$][$\{``s",``b"\},\{``\gamma"\}$]} is the $\CO_A^R$ relevant for $b\to s\gamma$.  
These effective operators are combined with their respective Wilson coefficients into the effective Hamiltonian:
\beq
H_{eff} = \sum_i C_i \CO_i 
\label{eq:Wilson}
\eeq
More details about the Wilson coefficients will be discussed in section \ref{sec:FFWilson}.

Next, \FFP\ loads \FFO\ and \FFW, the two central packages of \FF. The first contains all the observable functions, taking general Wilson coefficients as inputs. So in principle it could be used independently of the rest of \FF, for any model, provided one has a way to turn that model into Wilson coefficients. The second contains the code that compiles the analytic amplitudes and extracts the Wilson coefficients from them. We will describe them in more detail in sections \ref{sec:obs} and \ref{sec:wilson}.

Finally, \FFP\ defines the main function, also called \FF, that takes input spectra and turns them into flavor and CP observables. This function is called internally by the observable and constraint functions described in section \ref{sec:FFeval}.
\codebox{FormFlavor[VariableList]}{evaluate flavor observables for all compiled processes from the numerical input \code{VariableList} in the current active running mode\index{FormFlavor}}{0.6} 
\codebox{FormFlavor[VariableList,mode]}{as above, but for running mode \code{mode}}{0.55} \vspace{-6.5mm}
 \codebox{FFConstraints[FFout]}{compute the flavor constraints from the \code{FormFlavor} output in the current active running mode\index{FFConstraints}}{0.65}\vspace{-5.1mm}
 \codebox{FFConstraints[FFout,mode]}{as above, but for running mode \code{mode}}{0.55}

\noindent Here \code{VariableList} is the output from \code{CalcSpec} or \code{FFRunFile} (see \ref{sec:CalcSpec}) in the form presented in  \code{\$FFCompileVarList}. \code{FFConstraints} takes as input the output of the \code{FormFlavor} command, and outputs constraints in the form discussed in  table~\ref{tab:constraints} and section \ref{sec:FFeval}.

\section{The Observable Functions:  FFObservables}
\label{sec:obs}

In this section, we will walk through the explicit treatment of the many flavor and CP observables provided in the standard release of \FF. (Detailed instructions on how to add additional observables are discussed in section \ref{sec:newobs}.) These observables are listed in table \ref{tab:obs} and they are all loaded by the \FFO\ portion of the \FF\ package. We stress that \FFO\ is in principle a general-purpose, model-independent package in its own right, that takes Wilson coefficients at the new physics scale and returns flavor and CP observables. Even if the user does not take advantage of the rest of \FF's functionality, he or she may find \FFO\ useful on its own.

\begin{table}[tdp]
\begin{center}
\begin{tabular}{|c|c|c|}
\hline
Observable & Experiment & SM prediction \\
\hline \hline
$\Delta m_{K}$ &$ (3.484\pm0.006) \times 10^{-15}~\rm{GeV}~$ & $-$  \\
$\epsilon_K$ &$ (2.28\pm 0.011) \times 10^{-3}$ & $ (2.24\pm 0.19) \times 10^{-3}$~\cite{Ligeti:2016qpi} \\
\hline
$\Delta m_{B_d}$ &$ (3.36\pm0.02) \times 10^{-13}~\rm{GeV}$ & $ (4.21\pm0.34) \times 10^{-13}~\rm{GeV}$~\cite{Bazavov:2016nty}  \\
\hline
$\Delta m_{B_s}$ &$ (1.169\pm0.0014) \times 10^{-11}~\rm{GeV}$ & $ (1.303\pm0.078) \times 10^{-11}~\rm{GeV}$~\cite{Bazavov:2016nty} \\
\hline
$\Delta m_{D}$ &$ (6.2^{+2.7}_{-2.8}) \times 10^{-15}~\rm{GeV}$ & $-$ \\
\hline
$\text{BR}(K_L\to \pi^0 \nu \bar{\nu})$ & $ <  2.60 \times 10^{-8} $  (90\% CL) & $(2.49\pm 0.39)\times 10^{-11}$~\cite{Mescia:2007kn} \\
$\text{BR}(K^+\to \pi^+ \nu \bar{\nu})$ &$   (17\pm11) \times 10^{-11} $ & $(7.8\pm0.8)\times 10^{-11}$ ~\cite{Brod:2010hi} \\
\hline
$\text{BR}(B \to X_s \gamma) $ & $ (3.49\pm0.19)\times10^{-4}$ & $(3.36\pm0.23)\times10^{-4}$~\cite{Misiak:2015xwa}   \\
$A_{CP}(B \to X_s \gamma) $ & $ (1.5\pm2.0)\times10^{-2}$ & $(1.1\pm1.7)\times10^{-2}$~\cite{Benzke:2010tq}   \\
$\Delta A_{CP}(B \to X_s \gamma) $ & $ (5.0\pm4.2)\times10^{-2}$~\cite{Lees:2014uoa} & $(0.0\pm0.0)\times10^{-2}$~\cite{Benzke:2010tq}   \\
\hline
$\text{BR}(B \to X_d \gamma) $ & $ (1.41\pm0.57)\times10^{-5}$ \cite{delAmoSanchez:2010ae,Crivellin:2011ba} & $(1.54^{+0.26}_{-0.31})\times10^{-5}$~\cite{Crivellin:2011ba}   \\
\hline
$\text{BR}(B_s \to \mu^+ \mu^-)$ &$(2.9\pm0.7)\times 10^{-9}$~\cite{CMSandLHCbCollaborations:2013pla} & $(3.65\pm 0.23)\times 10^{-9}$~\cite{Bobeth:2013uxa}   \\
\hline
$\text{BR}(B_d \to \mu^+ \mu^-)$ &$ (3.6^{+1.6}_{-1.4}) \times 10^{-10}$~\cite{CMSandLHCbCollaborations:2013pla} & $(1.06\pm 0.09)\times 10^{-10}$~\cite{Bobeth:2013uxa}   \\
\hline
$|d_n|$ &$ <~2.9\times 10^{-26}~\rm{e~ cm}$   (90\% CL)& $\sim 10^{-34}\,\rm{e~cm}$~\cite{Czarnecki:1997bu} \\
\hline
$\text{BR}(\mu\to e\gamma)$ &$ <5.7\times10^{-13}$   (90\% CL) \cite{Adam:2013mnn} & $\sim 0$ \\
\hline
$\text{BR}(\tau\to\mu\gamma)$ &$ <4.4\times10^{-8}$  (90\% CL) & $\sim 0$ \\
\hline
$\text{BR}(\tau\to e\gamma)$ &$ <3.3\times10^{-8}$  (90\% CL)& $\sim 0$ \\
\hline
\end{tabular}
\end{center}
\caption{Current experimental and theoretical values used in \FF.   Except where noted, the experimental values are taken from recent PDG or HFAG fits~\cite{{Beringer:1900zz},{Amhis:2012bh}}.  No reliable theoretical prediction for $\Delta m_D$ currently exists.  Although literature on the subject exists, we do not use a theoretical prediction for $\Delta m_K$.\label{tab:obs} }
\index{Meson Mixing!$\Delta m_K$}
\index{Meson Mixing!$\Delta m_D$}
\index{Meson Mixing!$\Delta m_{B_d}$}
\index{Meson Mixing!$\Delta m_{B_s}$}
\index{Meson Mixing!$\epsilon_K$}
\index{Ktopinunu@BR$(K^+ \to \pi^+ \nu\bar\nu)$}
\index{Ktopinunu@BR$(K_L \to \pi^0 \nu\bar\nu)$}
\index{btosgamma@$b \to s \gamma$}
\index{btosgamma@$b \to s \gamma$!BR$(b \to X_s \gamma)$}
\index{btosgamma@$b \to s \gamma$!$A_{CP}$}
\index{btosgamma@$b \to s \gamma$!$\Delta A_{CP}$}
\index{btodgamma@$b \to d \gamma$}
\index{btodgamma@$b \to d \gamma$!BR$(b \to X_d \gamma)$}
\index{btodgamma@$b \to d \gamma$!BR$(b \to X_d \gamma)$}
\index{Bstomumu@$\text{BR}(B_s\rightarrow \mu^+ \mu^-)$}
\index{Bdtomumu@$\text{BR}(B_d\rightarrow \mu^+ \mu^-)$}
\index{Neutron EDM}
\index{mutoegamma@$\text{BR}(\mu\rightarrow e\gamma)$}
\index{tautoegamma@$\text{BR}(\tau\rightarrow e\gamma)$}
\index{tautomugamma\@$\text{BR}(\tau\rightarrow \mu\gamma)$}
\end{table}
   
Loading the \FFO\ package loads the individual observable module files. (There is one such file for each grouped set of rows in table \ref{tab:obs}.)  These are all located in the \file{Core/Observables} directory. Each observable module is a model-independent file containing an observable function,  as well as the current experimental values, and the experimental and theoretical uncertainties. We will describe the structure of the observable modules in more detail in the next subsection.

\FFO\ also defines the main function, also called \FFO. This function takes as input the RG scale of the Wilson coefficients, the effective Hamiltonian \code{Heff} as defined in (\ref{eq:Wilson}), and a list of desired observables. Then \FFO\ loops over these observables, calls the required observable function, and returns a list of the observables and their numerical values for the given Wilson coefficients.  
 \codebox{FFObservables[$\mu$,proclist,Heff]}{compute the flavor constraints for the processes contained in \code{proclist} (a subset of \code{\$FFProcessList}) for the effective Hamiltonian \code{Heff}  input at the scale $\mu$\index{FFObservables}}{0.52}

\subsection{Structure of the observable modules}

Within the observable module files, the first block of code provides some necessary linking to the model-dependent portion of the code.  

\begin{itemize}

\item Each observable module has an associated amplitude file containing analytic expressions for the Wilson coefficients used by the module.\footnote{Let us the stress the distinction between observables and amplitudes. An amplitude or process is an $S$-matrix calculation with a well-defined initial and final state,  for instance $\bar d s\to \bar s d$, from which Wilson coefficients are extracted.  An observable is something tied to a physical measurement, such as $\Delta m_K$ or BR$(B\to X_s\gamma)$.  Importantly, the map between observables and amplitudes is neither one-to-one nor onto. A single amplitude can give rise to multiple observables, for instance, the $KK$ mixing process contains the necessary information to evaluate both $\Delta m_K$ and $\epsilon_K$.  Meanwhile, a single observable can depend on multiple processes due to mixing under RG evolution, for instance BR$(B\to X_s\gamma)$ depends on both $b\to s\gamma$ and $b\to s g$.} The amplitude files are generated by the \CA\ package  (see section \ref{sec:CalcAmps} for details on this package), and they are stored within the \file{\{model\}/ObservableAmps} folder. 
(Within the \file{\{model\}/ObservableAmps/SubAmps} directory, the pre-compiled, built amplitudes are stored.)   The observable module ensures that the necessary amplitude file is loaded by appending it to the global \FF\ variable \code{\$FFAmpFileList}\index{\$FFAmpFileList}.

\item Each observable module inherits a name that is defined in the \CA\ amplitude file. The observable module  automatically finds the name from the amplitude file and appends it to  \code{\$FFProcessList}\index{\$FFProcessList}.

\item Lastly, it links the process name with the observable function name (defined in the final block of the file) via the line \blockc{\code{ObservableFunction[TempProcessName]=}\{\emph{observable function name}\}\code{;}\index{ObservableFunction}}
 where \code{TempProcessName} is a dummy variable storing the current process name.  This information is used by the main \code{FFObservables}\ routine.

\end{itemize}

Since each observable module can contain multiple observables (e.g.\ the K-K mixing observable module has both $\Delta m_K$ and $\epsilon_K$), the next block of code in the observable module steps through each observable, defines its \FF\ internal name \code{name}, and sets the parameters defining its current experimental status:
   \codebox{FFObsClass[name]}{observable's status (= 0, 1, or 2) (see Table \ref{tab:constraints})\index{FFObsClass}}{0.65} \vspace{-5.9mm}
   \codebox{FFExpValue[name]}{observable's current experimental value\index{FFExpValue}}{0.65} \vspace{-6.9mm}
   \codebox{FFExpUnc[name]}{observable's current experimental uncertainty\index{FFExpUnc}}{0.65}\vspace{-6.9mm}
   \codebox{FFSMValue[name]}{observable's current theoretical value (not used)\index{FFSMValue}}{0.65}\vspace{-6.9mm}
   \codebox{FFSMUnc[name]}{observable's current theoretical uncertainty\index{FFSMUnc}}{0.65}
The observable names are appended to the variable \code{\$FFObsNameList}\index{\$FFObsNameList} for later access, notably by the \code{FFConstraints} function.

Next within the code is an optional block that may contain global definitions, either functions or constant variables specific to the process, that are evaluated when \FF\ is loaded, rather than each time the process is called.  

The observable function is defined in the final block and contains information hardcoded from the literature on how to evaluate the  observables from the  high scale Wilson coefficients.   The input to each observable function is the effective Hamiltonian (\ref{eq:Wilson}), where $C_i$ are the numerical values of the Wilson coefficients (in general complex). The observable functions also need the scale where the Wilson coefficients were evaluated as an input. The output of each observable function is a list of related observables (e.g.\ $\Delta m_K$ and $\epsilon_K$) and their numerical values. 
  
For each observable function, there are two options that default to \code{True}: \code{IncludeSM}, which, when set to false, does not include the standard model contribution in the evaluation of the observable; and \code{QCDRG}, which when disabled stops the code from RG evolving the new physics contribution down to the lower scale, instead evaluating with the high scale Wilson coefficient value.  In some cases, such as the neutron EDM, and $\Delta m_D$, no standard model portion is included at all, so \code{IncludeSM} does nothing.  Similarly, many observables, such as $B_q\to\mu^+\mu^-$ and $\mu\to e\gamma$ have no QCD RG evolution, so \code{QCDRG} does nothing.  As these are primarily valuable for validation, it is recommended that users leave these settings set to \code{True}.\index{IncludeSM}\index{QCDRG}

We will now turn to a detailed description of each observable function.

\subsection{Meson Mixing}
\label{sec:MesonMixing}

 The matrix element relevant for meson mixing can be written as,
\bea
\frac{\langle X | H_{eff} | \bar X\rangle}{2m_X}  & \equiv M_{X,12} - \frac{i}2\Gamma_{X,12}\\
\frac{\langle \bar X | H_{eff} | X\rangle}{2m_X}  &= \, M^*_{X,12} - \frac{i}2\Gamma^*_{X,12}
 \label{eq:mXHeff}
\eea
where $m_X$ is the averaged meson mass, $H_{eff}$  is the effective Hamiltonian, and $M_{X,12}$ and $\Gamma_{X,12}$ are, respectively, the dispersive and absorptive parts of $\langle X | H_{eff} | \bar X\rangle$.   Short-distance effects contribute only to the dispersive portion in terms of $\abs{\Delta F}=2$ operators, whereas long-distance effects consisting of intermediate on-shell and off-shell particle exchanges, i.e., $(\abs{\Delta F}=1)^2$, can contribute to both  parts.  The relevant $\Delta F=2$ Wilson operators are shown in (\ref{eq:OpS})--(\ref{eq:OpT}) with $f_1=f_3=q_1$, and $f_2=f_4=q_2$,  for $(q_1,q_2)=(s,d),(c,u),(b,d),(b,s)$ for $K$, $D$, $B_d$ and $B_s$ respectively.  The short-distance portion of $M_{X,12}$ (which is all that is relevant for the $B_q$ systems and for new physics) is given by \cite{Buras:2001ra} 
\bea
M^{SD}_{X,12} &=\, \frac{m_X f_X^2}{24} \Bigg(
 8 B_V^{LL} (C_{V}^{LL}  + C_{V}^{RR}) -
 R_X \Bigg[ 4 B_{V}^{LR} C_{V}^{LR}  - 6 B_{S}^{LR} C_{S}^{LR}\Bigg] \\
 &\qquad  -R_X \Bigg[ 5 B_{S}^{LL} \lp C_{S}^{LL} + C_{S}^{RR}\rp + 12 B_{T}^{LL} \lp C_{T}^{LL} + C_{T}^{RR}\rp\Bigg] \Bigg)
 \label{eq:masterHeff}
\eea
where,
\beq
R_X \equiv \lp \frac{m_X}{m_{q_1}+m_{q_2}}\rp^2,
\eeq
and the non-perturbative $B$-parameters have been computed on the lattice.   For all mesons, only $C^{LL}_{V}$ is nonzero in the Standard Model.  The parameters relevant for meson mixing are summarized in Table \ref{tab:MesonProperties}.

 In terms of these quantities, the mass splitting mixing parameter can be expressed as \cite{Beringer:1900zz},
\beq
\Delta m_X= 2 \Re \left[ \lp M_{X,12} - \frac{i}2\Gamma_{X,12}\rp \sqrt{\frac{M^*_{X,12} - \frac{i}2\Gamma^*_{X,12}}{M_{X,12} - \frac{i}2\Gamma_{X,12}}} \right].
\label{DeltamXgen}\eeq
For both $X=B_d$ and $B_s$, where in the standard model the short-distance contributions are dominant, and  it is known experimentally that $\Gamma_{X,12}\ll M_{X,12}$, (\ref{DeltamXgen})  can be well approximated by,
\beq
\label{DeltamB}
\Delta m_X\approx 2\abs{M_{X,12}} = 2 \abs{ M^{SM}_{X,12} +M_{X,12}^{NP}}.
\index{Meson Mixing!$\Delta m_{B_d}$}\index{Meson Mixing!$\Delta m_{B_s}$}
\eeq 
For $X=K$ and $D$, long-distance contributions to $M_{X,12}$ are appreciable and even assumed to dominate in the case of $\Delta m_D$.   However, experimentally one has that $M_{X,12}/\Gamma_{X,12}$ is approximately real, and both $M_{X,12}$ and $\Gamma_{X,12}$ are predicted to be approximately real in the CKM convention used in \FF\ (where the CPV phase is primarily in $V_{td}$ and $V_{ub}$).  Using that information, one can approximate (\ref{DeltamXgen}) as,
\beq
\label{DeltamKD}
\Delta m_X\approx 2\,{\rm Re}\,M_{X,12} = \Delta m_{X,SM} +2\,{\rm Re}\,M_{X,12}^{NP}
\index{Meson Mixing!$\Delta m_K$}
\index{Meson Mixing!$\Delta m_D$}
\eeq
 In the case of $\Delta m_D$, this approximation is not great, however, relative to the enormous theoretical uncertainties in $\Delta m_D$, this treatment is sufficient to determine regions where new physics contributions overwhelm the experimental measurement.

\begin{table}[tpd]
\begin{center}
\small
\begin{tabular}{|c|ccc|ccccc|c|}
\hline
\!Meson\! & \!\!$m_X$\footnotesize(GeV)\normalsize\!\! & \!\!$f_X$\footnotesize(GeV)\normalsize\!\! & $R_X$ & $B^{LL}_V$ & $B_{V}^{LR} $ & $B_{S}^{LR} $ & $B^{LL}_{S} $ & $B^{LL}_{T}$  & $C^{LL}_{V,SM}|_{\mu=m_b}$(GeV$^{-2}$)\normalsize  \\
\hline 
$K$ & $0.4976$ & $0.160$ & $24.3$ & $0.56$ & $0.85 $ & $1.08$ & $0.62$ & $0.43$ & $-$ \\
$D$ &  $1.8645$ & $0.209$ & $3.20$ & $0.76$ & $0.97$ & $0.95$ & $0.64$ & $0.39$ &  $-$    \\ 
${B_d}$ & $5.2796$ & $0.188$ & $1.65$ &  $0.913$ & \!$1.838 $\! & \!$1.145$\! & \!$0.761$\! & \!$0.555$\! &  $(2.34-2.20i) \!\times\! 10^{-12}$  \\
${B_s}$ & $5.3668$ & $0.226$ & $1.65$ &  $0.952$ & \!$1.799 $\! & \!$1.125$\! & \!$0.806$\! &\!$0.610$\! &  $(6.96-0.26i) \!\times\! 10^{-11}$ \\
\hline
\end{tabular}
\normalsize
\caption{ Parameters of mesons relevant for the \FF\ meson mixing observables.   For both the $K$ and $D$ system, $f_X$ and $B_{V,X}^{LL}$ are taken from the FLAG review \cite{Aoki:2013ldr}, while  $f_{B_d}$ and $f_{B_s}$ come from \cite{Dowdall:2013tga} and \cite{Rosner:2015wva}, respectively.  The other non-perturbative $B$-parameters were collected from several sources: for $\Delta m_K$ from \cite{Bae:2013mja} (at $\mu=2$ GeV), $\Delta m_D$ from \cite{Carrasco:2013jaa} (at $\mu=3$ GeV, rescaled for a common $R_D$, and converted to our basis using $B^{LL}_T=\frac 53 B_2-\frac 23 B_3$), and for $B_s$ and $B_d$ from \cite{Bazavov:2016nty} (at $\mu=\bar{m_b}=4.18$ GeV, rescaled to match the form in (\ref{eq:masterHeff}), and converted to our basis).  $C^{LL}_{V,SM}|_{\mu=m_b}$ are the \FF\ values in the CKM basis of the PDG \cite{Beringer:1900zz}. \label{tab:MesonProperties}}
\index{Meson Mixing}
\end{center}
\end{table}

The QCD RG evolution of the Wilson coefficients in (\ref{eq:masterHeff}) is controlled by the function \code{DeltaF2RG}.  This RG evolution \cite{Buras:2001ra}  is for Wilson coefficients in the NDR-$\bar{\mbox{MS}}$ with the BMU evanescent operator scheme \cite{Buras:2000if}.

\codebox{DeltaF2RG[$\mu_{L}$,$\mu_{H}$,list]}{RG evolve $\abs{\Delta F}\!=\!2$ Wilson coefficients from high scale, $\mu_H$, to low scale, $\mu_L$, with input and output in the form \\ \code{list}$=\{C_V^{LL},C_V^{RR},C_V^{LR},C_S^{LR},C_S^{LL},C_T^{LL},C_S^{RR},C_T^{RR}\}$\index{DeltaF2RG}}{0.65} \vspace{1mm}

The indirect $CP$-violation in $K^0 - \bar K^0$ mixing, $\epsilon_K$, can be expressed as \cite{Buras:2015jaq}:
\beq
\epsilon_K \equiv \frac{\tilde \kappa_\epsilon \, {\rm Im}\,M_{X,12}}{\sqrt 2 \Delta m_{K,exp}} = \epsilon_{K,SM} + \frac{\tilde \kappa_\epsilon \, {\rm Im}\,M_{X,12}^{NP}}{\sqrt 2 \Delta m_{K,exp}}
\label{eq:epsK}
\index{Meson Mixing!$\epsilon_K$}
\eeq
where $\tilde \kappa_\epsilon =0.94\pm0.02$ accounts for small long-distance corrections in $\Im \Gamma_{12}$ and $\Im M_{12}$ from $\lp H_{\abs{\Delta F}=1}\rp^2$ contributions \cite{Buras:2010pza}. Although we refer to this here and in the manual and code as $\epsilon_K$, many references in the literature include a factor of the ``superweak phase" which has been experimentally determined to be very close to $\pi/4$ \cite{Beringer:1900zz}.  New short distance physics will not change the phase of $\epsilon_K$ appreciably, so for the treatment of this observable, it makes sense to remove this phase, as we have done in (\ref{eq:epsK}).

 The SM contribution for $\Delta m_{B_d,B_s}$ is hardcoded and combined as in (\ref{DeltamKD}); whereas no SM contribution is used for $\Delta m_K$ and $\Delta m_D$ due to the large uncertainties \cite{Brod:2011ty}.  The following \FF\ functions evaluate these meson mixing observables:

\codebox{KKmixing[$\mu$,Heff]}{evaluate both $\Delta m_K$ and $\epsilon_K$ with Wilson coefficients input at the scale $\mu$\index{KKmixing}}{0.68}\vspace{-4.9mm}
\codebox{DDmixing[$\mu$,Heff]}{evaluate $\Delta m_D$ with Wilson coefficients at the scale $\mu$\index{DDmixing}}{0.68}\vspace{-6.7mm}
\codebox{BdBdmixing[$\mu$,Heff]}{evaluate $\Delta m_{B_d}$ with Wilson coefficients at the scale $\mu$\index{BdBdmixing}}{0.66}\vspace{-6.7mm}
\codebox{BsBsmixing[$\mu$,Heff]}{evaluate $\Delta m_{B_s}$ with Wilson coefficients at the scale $\mu$\index{BsBsmixing}}{0.66}

\subsection{$b \to q \gamma$}
\label{sec:btoqgamma}

The $b \to s \gamma$ and $b \to d \gamma$ observables come primarily from dimension 5 effective Hamiltonian (\ref{eq:OpA})--(\ref{eq:OpG}) with $f_2=b$ and $f_1=s$ or $d$. The leading order branching ratio can be written as,
\be
[\text{BR}(b\rightarrow q \gamma)]_{LO}=  c_\gamma v^2 \Big( \abs{ C_A^R}^2 + \abs{  C_A^L}^2 \Big)
\label{eq:bsgsimple}
\ee
where \cite{Bauer:2004ve,Czakon:2015exa}
\bea
 c_\gamma  &=\; \lp8 \pi^2 \rp^2\frac{6}{\pi} \frac{\text{BR}(b\rightarrow X_c e \nu)_{\text{\tiny{EXP}}}}{\Phi \abs{V_{cb} }^2} \frac{v^2}{m_b^2}\alpha_{\text{\tiny{EM}}} =  3.3 \times 10^{7}\\
\Phi &=\; \abs{\frac{V_{ub}}{V_{cb}}}^2 \frac{\text{BR}(b\rightarrow X_c e \nu)}{\text{BR}(b\rightarrow X_u e \nu)} = 0.569.
\eea
The branching ratio into charm final states is used to remove the hadronic matrix element, and the $\Phi$ factor  accounts for the nontrivial phase space differences between the two decays (due mostly to the charm quark mass).  However, this leading order treatment is insufficient for an accurate SM prediction of $b\to s \gamma$.   Currently, the uncertainty from the combined experimental world average and from the most precise theory determination are very close in size.  As BR$(B\to X_s \gamma)$ is often one of the most constraining observables for BSM physics, an accurate SM prediction is essential.  In this section, we explain how \FF\ handles the various $b \to q \gamma$ observables.  Due to the expansive literature on the subject, we will not replicate many expressions and refer interested users to the primary references.  

Both $\text{BR}(B\rightarrow X_s \gamma)$ and $\text{BR}(B\rightarrow  X_d \gamma)$ can be expressed as,
\beq
\text{BR}(B\rightarrow X_q \gamma)_{E>E_0}= c_\gamma \frac{\abs{V_{tq}^*V_{tb}}^2m_b^2}{ \lp8 \pi^2 \rp^2 v^2}  \bigg( P(E_0) + N(E_0)\bigg),
\label{eq:bqgmaster}
\index{btosgamma@$b \to s \gamma$}
\index{btodgamma@$b \to d \gamma$}
\eeq
where $P(E_0)$ and $N(E_0)$ are the perturbative and small, non-perturbative contributions, respectively.  The leading order expression for $P(E_0)$ can be easily extracted from (\ref{eq:bsgsimple})--(\ref{eq:bqgmaster}).  As $\text{BR}(b\rightarrow s \gamma)$ and $\text{BR}(b\rightarrow d \gamma)$ are known experimentally to very different levels of precision, in \FF\ we give a slightly different treatment for each of these observables.  We will first discuss the $b\to s\gamma$ process and the three associated observables,  before discussing the treatment of $b\to d\gamma$.
 
 \begin{table}[tpd]
\begin{center}
\small
\begin{tabular}{|c|rrrrrr|rr|}
\hline
$C_{eff}(\mu_b)$& 1&2&3&4&5&6&7&8 \normalsize  \\
\hline 
$C^{(0)}_{eff}$& $-0.8999$&$1.073$&$-0.0151$&$-0.1393$&$0.0014$&$0.0032$&$-0.3848$&$-0.1776$ \\
$C^{(1)}_{eff}$& $14.94$&$-2.210$&$0.0842$&$-0.5902$&$-0.0207$&$-0.0069$&$2.087$&$-0.6311$ \\
$C^{(2)}_{eff}$& &&&&&&$18.86$& \\
\hline
\end{tabular}
\normalsize
\caption{The standard model effective Wilson coefficients relevant for BR$(B\to X_s\gamma)$ evaluated at $\mu_b =2$ GeV.  At second order in $\alpha_s$, only $C^{(2)}_{eff,7}$ is necessary. \label{tab:bsgCeff}} \vspace{-6mm}
\end{center}
\end{table}

To NNLO order, the perturbative portion of BR$(B\to X_s \gamma)$ can be written \cite{Misiak:2006ab}, 
\beq
P(E_0) = \sum_{i,j=1}^8 C_{eff,i}(\mu_b) K_{ij}(\mu_b,E_0) C_{eff,j}^*(\mu_b)
\eeq
 where $K_{ij}(\mu_b,E_0)$ is a Hermitian matrix, and the effective Wilson coefficients $C_{eff,i}$ are evaluated at the scale $\mu_b$.  The values of $C_{eff,i}^{(1,2)}$ and $K_{ij}^{(1,2)}$ have been determined through decades of work \cite{Czakon:2015exa,Misiak:2006ab,Buchalla:1997kz,Bobeth:1999mk,Gambino:2001ew,Bieri:2003ue,Huber:2005ig,Asatrian:2006rq,Czakon:2006ss,Boughezal:2007ny,Ewerth:2008nv,Asatrian:2010rq,Ferroglia:2010xe,Misiak:2010tk,Kaminski:2012eb,Huber:2014nna}.   For details of the definitions of $C_{eff,i}$, see \cite{Czakon:2015exa} and references within.   For our purposes, it suffices to know that  $C_{eff,7}\sim C_A^R$ and $C_{eff,8}\sim C_G^R$,\footnote{In actuality, $C_{eff,7}=y_0 C_A^R+\sum_i^6 y_i C_{i}$ and $C_{eff,8}=z_0 C_G^R+\sum_i^6 z_i C_{i}$ \cite{Misiak:2006ab}.  These details are not important for the new physics, as the leading order contributions to $C_{1-6}$ are tree level, while $C_{7,8}$ leading order contributions enter at one-loop level, so new physics added at loop level in \FF\ only contributes to  $C_{7}$ and $C_{8}$.   That  $C_{eff,7}$ etc are \emph{linear} in the other $C_i$ Wilson coefficients means that the SM and new physics contributions can be completely decoupled.}  
  whereas the others are coefficients of four Fermi or $(\bar b s)\{current\}$ operators. Both $C_{eff,i}$ and $K_{ij}(\mu_b,E_0)$ can be expanded perturbatively in powers of $\tilde \alpha_s = \frac{\alpha_s(\mu_b)}{4\pi}$. The values for $C_{eff,i}$ in the SM up to second order in $\tilde\alpha_s$ are given in table \ref{tab:bsgCeff}; while $K^{(0)}=\delta_{i7}\delta_{j7}$ and  the values for $K_{ij}^{(1,2)}$ (at the scale $\mu_b=2$~GeV and $E_0=1.6$~GeV) are  \cite{Misiak:2006ab}:
   \beq
K^{(1)}=
 \lp \!\!
 \begin{tabular}{cccccc|cc}
0.0029\!\!&\!\! -0.017\!\!&\!\! 0.0005\!\!&\!\! 0.0042\!\!&\!\! 0.0054\!\!&\!\! -0.062\!\!&\!\! 0.091\!\!&\!\! -0.0017\\
 -0.017\!\!&\!\! 0.104\!\!&\!\! -0.0029\!\!&\!\! -0.026\!\!&\!\! -0.032\!\!&\!\! 0.37\!\!&\!\! -0.55\!\!&\!\! 0.010\\ 
 0.0005\!\!&\!\! -0.0029\!\!&\!\! 0.029\!\!&\!\! -0.0048\!\!&\!\! 0.32\!\!&\!\! -0.028\!\!&\!\! 8.37\!\!&\!\! -0.055\\
  0.0042\!\!&\!\! -0.026\!\!&\!\! -0.0048\!\!&\!\! 0.0008\!\!&\!\! -0.053\!\!&\!\! 0.0047\!\!&\!\! -1.73\!\!&\!\! -0.204\\ 
  0.0054\!\!&\!\! -0.032\!\!&\!\! 0.32\!\!&\!\! -0.053\!\!&\!\! 3.55\!\!&\!\! -0.084\!\!&\!\! 123.5\!\!&\!\! -0.838\\ 
  -0.062\!\!&\!\! 0.37\!\!&\!\! -0.028\!\!&\!\! 0.0047\!\!&\!\! -0.084\!\!&\!\! 1.81\!\!&\!\! 18.36\!\!&\!\! -1.93\\ 
  \hline
  0.091\!\!&\!\! -0.55\!\!&\!\! 8.37\!\!&\!\! -1.73\!\!&\!\! 123.5\!\!&\!\! 18.36\!\!&\!\! 5.62\!\!&\!\! -0.507\\ 
  -0.0017\!\!&\!\! 0.010\!\!&\!\! -0.055\!\!&\!\! -0.204\!\!&\!\! -0.838\!\!&\!\! -1.93\!\!&\!\! -0.507\!\!&\!\! 0.452
   \end{tabular}\!\!
  \rp
    \label{eq:K1mat}
 \eeq
 and \cite{Czakon:2015exa}
  \beq
K^{(2)}=
 \lp \!\!
 \begin{tabular}{cccccc|cc}
0.11&-0.69&0&0&0&0& 9.11&0.22\\
-0.69& 4.12&0&0&0&0& -8.86&-1.29\\
0&0&0&0&0&0&0&0\\
0&0&0&0&0&0&0&0\\
0&0&0&0&0&0&0&0\\
0&0&0&0&0&0&0&0\\
  \hline
9.11& -8.86&0&0&0&0&-37.32&-13.41\\
0.22&-1.29&0&0&0&0&-13.41& 22.32
   \end{tabular}\!\!
  \rp.
  \label{eq:K2mat}
 \eeq
 The terms in $K^{(2)}$ aligned with $C_{eff,3-6}$ are believed to be very small, and thus approximated to be 0 at this level.   The small non-perturbative correction of $N(E_0)=3.82\times 10^{-7}$ \cite{Ewerth:2009yr,Benzke:2010js,Alberti:2013kxa} is included in \FF, but has almost no effect on the net BR.\footnote{We are extremely grateful to M.~Misiak for providing code to compute the standard model $C_{eff}^{(i)}$, $K_{ij}^{(1,2)}$ and $N(E_0)$ values used in \FF.}  

In order to include new physics in the $b\to s \gamma$ branching ratio, the NP contribution is RG evolved from the high scale, down to the low scale using the \FF\ function \code{btosgammaRG}.  
\codebox{btosgammaRG[$\mu_{L}$,$\mu_{H}$,list]}{RG evolve dimension 5 Wilson coefficients from high scale, $\mu_H$, to low scale, $\mu_L$, with input and output in the form \code{list}$=\{C_A^{R},C_G^{R},C_A^{L},C_G^{L}\}$\index{btosgammaRG}}{0.62} \vspace{-1mm}
\noindent This evolution only includes mixing between $C_7$ $(C_A^R)$ and $C_8$ $(C_G^R)$  and between $\tilde C_7$  $(C_A^L)$ and $\tilde C_8$ $(C_G^L)$.   Importantly, the RG evolution of the dimension-6 Wilson coefficients, $C_{1-6}$, is not affected by the dimension-5  coefficients, $C_{7,8}$; and $C_{1-6}$ contribute to the RG evolution of $C_{7,8}$ \emph{linearly}. Since we do not include new physics in the dimension-6 operators (which would contribute to $b\to q\gamma$ at a higher order), the effects of $C_{1-6}$, including their mixing through the RG, have already been included in the SM portion.  

It was realized long ago that direct CP asymmetries observed in $b\to s\gamma$ could be a sign of new physics \cite{Soares:1991te}, and some early estimates predicted very small uncertainty ($\pm 0.5\%$) in the SM \cite{Hurth:2003dk}.  However, initially neglected long-distance contributions to the asymmetry greatly increase the uncertainty \cite{Benzke:2010tq}.  Still, a model predicting a substantially larger than measured value of $A_{CP}(B \to X_s \gamma)$ could be conclusively ruled out.  The direct CP symmetry in $B$ decays is defined as:
\beq
A_{CP}(B \to X_s \gamma) \equiv \frac{\Gamma\lp \bar B \to X_{s} \gamma \rp-\Gamma\lp  B \to X_{\bar s} \gamma \rp}{\Gamma\lp \bar B \to X_{s} \gamma \rp+ \Gamma\lp  B \to X_{\bar s} \gamma \rp},
\eeq
  this can be expressed as \cite{Benzke:2010tq}, 
   \beq
 A_{CP}\big|_{E_\gamma>E_0} =  A_{CP}^{dir}\big|_{E_\gamma>E_0} +A_{CP}^{res}\big|_{E_\gamma>E_0} 
\eeq
 where $A_{CP}^{dir}$ and $A_{CP}^{res}$ are the direct CPV contribution and the resolved contribution that accounts for the hadronic substructure of the photon \cite{Lee:2006wn,Benzke:2010js}.    The direct contribution can be approximated with \cite{Kagan:1998bh,Ali:1998rr,Benzke:2010tq},
 \beq
 A_{CP}^{dir}\big|_{E_\gamma>E_0}\approx \alpha_s \lp \frac{40}{81}\Im\left[\frac{C_2}{C_7} \right] -\frac49\Im\left[\frac{C_8}{C_7} \right]  - \frac{40\Lambda_c}{9m_b}\Im\left[\lp1+\epsilon_s\rp\frac{C_2}{C_7} \right] \rp
\index{btosgamma@$b \to s \gamma$!$A_{CP}$}
\eeq
 where $\epsilon_s=\frac{V_{ub}V_{us}^*}{V_{tb}V_{ts}^*}$, corrections of $\order{\frac{\Lambda_{QCD}^2}{m_b^2}}$ have been dropped, and
 \beq
 \Lambda_c\equiv \frac{m_c^2}{m_b} \lp 1 -\frac25 \ln \frac{m_b}{m_c}+\frac45 \ln^2 \frac{m_b}{m_c} -\frac{\pi^2}{15} \rp.
\eeq
 The resolved photon contribution can be expressed as \cite{Benzke:2010tq}, 
 \beq
  A_{CP}^{res}\big|_{E_\gamma>E_0}\approx \frac{\pi}{m_b} \lp \Im\left[\lp1+\epsilon_s\rp\frac{C_2}{C_7} \right] \tilde \Lambda^c_{27}  -\Im\left[\epsilon_s \frac{C_2}{C_7} \right]\tilde \Lambda^u_{27} + 4\pi\alpha_s \Im\left[\frac{C_8}{C_7} \right] \tilde \Lambda^{\bar B}_{78}  \rp.
 \eeq
The hadronic $\tilde\Lambda$ parameters are extremely uncertain, but will be estimated to their central values of $\tilde \Lambda_{27}^c=0.001$ GeV, $\tilde \Lambda_{27}^u=0.1$ GeV, and $\tilde \Lambda^{\bar B}_{78}$  additionally depends linearly on the charge of the spectator quark.  We will use that BR$[\Upsilon(4S)\to B^+B^- \,(B^0 \bar B^0)]=0.513\,(0.487)$ to write $\tilde \Lambda^{\bar B}_{78} = -(2\times0.513-0.487)/3 \tilde \Lambda_{78} =- 0.18 \tilde \Lambda_{78}$, with $\tilde \Lambda_{78}\approx 0.1$ GeV.  In \FF, we apply an SM theory uncertainty of $\pm1.7\%$ \cite{Benzke:2010tq} to this observable.   

It was also pointed out in reference \cite{Benzke:2010tq} that a resolved CP asymmetry difference between charged and neutral  meson decays can be a sensitive probe of new physics.  This quantity, 
\bea
\Delta A_{CP}(B \to X_s \gamma)\big|_{E_\gamma>E_0} \equiv& \lp A_{CP}(B \to X_s^- \gamma)-A_{CP}(B \to X_s^0 \gamma)\rp \big|_{E_\gamma>E_0} \\ \approx& \; 4 \pi^2 \alpha_s \frac{\tilde \Lambda_{78}}{m_b}\Im\left[\frac{C_G^R}{C_A^R} \right] 
\index{btosgamma@$b \to s \gamma$!$\Delta A_{CP}$}
\eea
 where the hadronic parameter $\tilde \Lambda_{78}$ is the dominant source of uncertainty, finishes in the standard model and serves as an additional constraint on new physics.  In \FF,  we follow \cite{Benzke:2010tq} and use central value  $\tilde \Lambda_{78}=0.1$ GeV, although more recent QCD sum rule calculations \cite{Nishikawa:2011qk} suggest a larger value such as $0.3$ GeV might be more accurate.  For excluding parameter points, a smaller value is more conservative.
 
The branching ratio and both CP asymmetries in $b\to s\gamma$ are evaluated with the \FF\ function \code{btosgamma}.
\codebox{btosgamma[$\mu$,Heff]}{evaluate BR$(B \!\rightarrow \! X_s \gamma)$, $A_{CP}(B\!  \to\!  X_s \gamma) $, and $\Delta A_{CP}(B\!\to\!X_s \gamma) $  with Wilson coefficients input at the scale $\mu$\index{btosgamma}}{0.68}

Finally, for $\text{BR}(B \rightarrow X_d \gamma)$, which is currently known to a much lower experimental precision, we follow the treatment in \cite{Hurth:2003dk} to derive the SM branching ratio.  Unlike in the case of $\text{BR}(B\rightarrow X_s \gamma)$ this result uses Wilson coefficients evaluated at the scale $m_t$ as input.  As the treatment uses Wilson coefficients evaluated at $m_t$, disabling \code{QCDRG} only removes the running of new physics Wilson coefficients from $m_{SUSY}$ to $m_t$.  We can express $P(E_0)$ in (\ref{eq:bqgmaster}) as \cite{Hurth:2003dk},
\bea
P(E_0)=& \; a_0 +a_{AA}\lp \abs{R_A}^2+\abs{\tilde R_A}^2 \rp+ a_{GG}\lp \abs{R_G}^2+\abs{\tilde R_G}^2 \rp +a^r_A \Re R_A+a^i_A \Im R_A\\
 &  +a^r_G \Re R_G+a^i_G\Im R_G + a_{\epsilon\epsilon}\abs{\epsilon_d}^2 +a^r_\epsilon \Re \epsilon_d +a^i_\epsilon \Im \epsilon_d \\
&+a^r_{A\epsilon}\Re\lp R_A \epsilon^*_d\rp +a^i_{A\epsilon}\Im\lp R_A \epsilon^*_d\rp +a^r_{G\epsilon}\Re\lp R_G \epsilon^*_d\rp +a^i_{G\epsilon}\Im\lp R_G \epsilon^*_d\rp \\
&+a^r_{GA}\Re\lp R_G R_A^*+ \tilde R_G \tilde R_A^* \rp +a^i_{GA}\Im\lp R_G R_A^*+ \tilde R_G \tilde R_A^* \rp
\label{eq:PE0bdg}
\index{btodgamma@$b \to d \gamma$}
\eea
where,
\beq
\epsilon_d=\frac{V_{ub}V_{ud}^*}{V_{tb}V_{td}^*} , \qq R_X = \frac{C_X^R(\mu=m_t)}{C_{X,SM}^R(\mu=m_t)}, \qq \tilde R_X = \frac{C_X^L(\mu=m_t)}{C_{X,SM}^R(\mu=m_t)},
\eeq
and the $a$ coefficients are given in table \ref{tab:bsdai}.

 \begin{table}[tpd]
\begin{center}
\small
\begin{tabular}{|cc|cc|cc|cc|}
\hline
$a_0$ & 6.9120 & 
$a_{AA}$ & 0.8161&
$a^r_A$&4.5689 &
$a^i_A$ & 0.2167 \\
$a_{GG}$ & 0.0197 & 
$a^r_G$ &0.5463 &
$a^i_G$ & $-0.1105$ &
$a_{\epsilon\epsilon}$ & 0.3787 \\
$a^r_\epsilon$ & $-2.6679$ &
 $a^i_\epsilon$ & 2.8956 &
$a^r_{GA}$ &0.1923 & 
$a^i_{GA}$ &$-0.0487$ \\
$a^r_{\epsilon A}$ &$-1.0940 $  &
$a^i_{\epsilon A}$ &$-1.0447$ &
$a^r_{\epsilon G}$ &$-0.0819 $ & 
$a^i_{\epsilon G}$ &$-0.0779$ \\
\hline
\end{tabular}
\normalsize
\caption{Numerical values for the coefficients in $P(E_0)$ (\ref{eq:PE0bdg}) relevant for BR$(B\to X_d\gamma)$.  Values shown are for $E_0=1.6$ GeV and $m_c/m_b=0.29$ \cite{Hurth:2003dk}. \label{tab:bsdai}}
\end{center}
\end{table}

BR$(B\to X_d\gamma)$ is evaluated with the \FF\ function \code{btodgamma}. 
\codebox{btodgamma[$\mu$,Heff]}{evaluate BR$(B\to X_d \gamma)$ with Wilson coefficients input at the scale $\mu$\index{btodgamma}}{0.68}

\subsection{$K\to \pi \nu\nu$}
\label{sec:Ktopinunu}

 Rare $K\to \pi \nu\nu$ decays can be expressed in terms of the Wilson operators $C^{ML}_{V,\ell_1\ell_2}$ where we have $f_1=d$, $f_2=s$, $f_3=\nu_{\ell_1}$, $f_4=\nu_{\ell_2}$ in (\ref{eq:OpV}).  The branching ratios for charged and neutral $K\rightarrow\pi\nu\bar\nu$ can be written \cite{Buras:2015qea}, 
\bea
\label{eq:Kpinunucomp}
\text{BR}(K^\pm\rightarrow\pi^\pm\nu\bar\nu)=&\;\frac{ c_+}{3} v^4 (1+\Delta_{EM}) \sum_{\ell_1,\ell_2=e,\mu,\tau}\left| {C}^{LL}_{V, \ell_1\ell_2}+ {C}^{RL}_{V,\ell_1\ell_2}\right|^2  \\
 \text{BR}(K_L\rightarrow \pi^0 \nu \bar\nu)=&\;\frac{ c_0}{3} v^4(1+\delta_\epsilon) \sum_{\ell_1,\ell_2=e,\mu,\tau} \lp \Im\!\left[  {C}^{LL}_{V, \ell_1\ell_2} + {C}^{RL}_{V, \ell_1\ell_2} \right] \rp^2
\eea 
where $v=246 \gev$, $\delta_\epsilon\approx - 0.011$ \cite{Buchalla:1996fp} accounts for the indirect CP violation in mixing, $\Delta_{EM}= -0.003$ \cite{Buras:2015qea} accounts for electromagnetic effects, and 
\bea
 c_+ &= \; \frac{3 \, r_{K^+} }{2 \abs{ V_{us}}^2} \, \text{BR}(K^\pm\rightarrow \pi^0 e^\pm \nu) = 1.35 \\
c_0 &= \; \frac{3 \, r_{K^0}}{2 \abs{ V_{us}}^2} \frac{\Gamma(K^\pm)}{ \Gamma(K_L)}\, \text{BR}(K^\pm\rightarrow \pi^0 e^\pm \nu) = 5.84. 
\eea
Here, the branching ratio (0.0507) has been included to remove dependence on the hadronic matrix element, and factors $r_{K^+}=0.901$ and $r_{K^0}=0.944$  contains  electroweak corrections and isospin violating quark mass effects that were computed in \cite{Marciano:1996wy}.   The RG evolution of the relevant $C^{ML}_{V,\ell_1\ell_2}$ operators is negligible \cite{Buras:1998raa}.

In the standard model, lepton number is a good symmetry, so all mixed flavor operators ($\ell_1\neq \ell_2$) vanish.  Meanwhile, the top loop contributions have no sensitivity to different lepton generations, whereas the charm loop contributions are the same for $e$ and $\mu$, but different for $\tau$.   Thus, the SM contribution can be expressed as,
\beq
({C}^{LL}_{V,\ell\ell})_{SM} =  \frac{\alpha_2 }{\pi v^2} \lp \lambda_c X^\ell_c + \lambda_t  X_t \rp 
\eeq
where $\lambda_i =  V^*_{is}V_{id}$, $X_t =1.481$ \cite{Buras:2015qea}, $X^e_c=X^\mu_c=1.055\times10^{-3}$, and $X^\tau_c=7.01\times10^{-4}$  \cite{Buras:1998raa}.

As for the new physics contribution, although lepton flavor violation is treated in \FF, the effect of lepton flavor violation on this observable in most models should be small, so the current implementation of $K\to \pi \nu\nu$ does not treat this possibility.  We will simplify expressions by further assuming lepton flavor universality, i.e. $C^{XY}_{V,NP,\ell_1\ell_2}\equiv C^{XY}_{V,NP}\delta_{\ell_1\ell_2}$.   We can then make a simplifying approximation to equation (\ref{eq:Kpinunucomp}) yielding,
\bea
\text{BR}(K^\pm\rightarrow\pi^\pm\nu\bar\nu)\approx&\;  c_+  (1+\Delta_{EM})  v^4 \left| {C}^{LL}_V+ {C}^{RL}_V\right|^2  \\
\text{BR}(K_L\rightarrow\pi^0\nu\bar\nu)\approx&\;  c_0  \lp 1+\delta_{\epsilon} \rp v^4 \Im\! \left[{C}^{LL}_V+ {C}^{RL}_V\right]^2 
\index{Ktopinunu@BR$(K^+ \to \pi^+ \nu\bar\nu)$}
\index{Ktopinunu@BR$(K_L \to \pi^0 \nu\bar\nu)$}
\eea
where $C^{XY}_V=C^{XY}_{V,NP}+C^{XY}_{V,SM}$, with 
\bea
{C}^{RL}_{V,SM} &=\; 0 \\
{C}^{LL}_{V,SM} &=\;  \frac{\alpha_2 }{\pi v^2} \lp \lambda_c \lp P_c+\delta P_{c,u}\rp + \lambda_t  X_t \rp =(-12.5+3.7 i)\times 10^{-11}\gev^{-2},
\eea
and $P_c$ is now the charm loop contribution averaged over the three neutrino flavors,
\bea
P_c &= \; \lp \frac23 X^e_{NL} +\frac13 X^\tau_{NL} \rp \sim 9.37 \times 10^{-4}.
\eea
  Using $P_c$ simplifies the expression at the cost of reducing the standard model charm contribution by roughly 3\%.  This  corresponds to a 0.3\% decrease of the overall SM contribution, which, compared to the current theoretical uncertainty, is completely negligible. Importantly, interference  between standard model and new physics contributions are properly captured with this simplification.  Contained in $\delta P_{c,u}$ are long-distance and dimension-8 contributions which increase the effective $P_c$ by about $10\%$ \cite{Isidori:2005xm},
\bea
\delta P_{c,u} &= \; \frac{\pi^2 f_\pi^2}{m_W^2}\lp \frac{4\abs{G_8}}{\sqrt2 G_F} -\frac43 \rp \sim \lp 1.08 \pm 0.54\rp \times 10^{-4}.
\eea
 For $K_L\rightarrow\pi^0\nu\bar\nu$, the $\lambda_t  X_t$ portion of the standard model dominates as $\lambda_c P_c$ is approximately real. 
 
 We note that the \FF\ values are above the ``official" SM predictions shown in table \ref{tab:obs}.  This is largely due to the change in central value of $\abs{V_{cb}}=A\lambda^2$ from $0.0406 \to 0.0418$ (between \cite{Brod:2010hi} and \cite{Charles:2004jd,CKMfitter}).  As $\lambda_t^2\propto A^4$ and $\lambda$ has not changed appreciably, this is an $\order{10\%}$ enhancement to both $K\to \pi \nu\nu$ branching ratios.  This shift is of course entirely consistent with \cite{Brod:2010hi}, where it was clearly presented that the dominant uncertainty was in $\abs{V_{cb}}$.  This is further supported by the larger values claimed in \cite{Buras:2015qea}.
  
Both $K\to\pi\nu\nu$ decays are evaluated with the \FF\ function \code{Ktopinunu}.
\codebox{Ktopinunu[$\mu$,Heff]}{evaluate both BR$(K_L\rightarrow\pi^0\nu\bar\nu)$ and BR$(K^\pm\rightarrow\pi^\pm\nu\bar\nu)$ with Wilson coefficients input at the scale $\mu$\index{Ktopinunu}}{0.68}

\subsection{$B_q\to \mu^+\mu^-$}
\label{sec:Bqtomumu}

The $B_q\to \mu^+\mu^-$ observables can be expressed in terms of  operators within (\ref{eq:OpS})--(\ref{eq:OpV}) for $f_1=b$, $f_2=s,d$ and $f_3=f_4=\mu$. The branching ratio for $B_{s,d}\to\mu^+\mu^-$ can be written as \cite{Bobeth:2002ch,Dedes:2008iw}:
\begin{equation}
\text{BR}(B_i\rightarrow \mu^+ \mu^-) = X_i \left[  \left(1-\frac{4 m_\mu^2}{m_{B_i}^2}\right) | {F}^{(i)}_S |^2 + | {F}^{(i)}_P + {F}^{(i)}_A |^2\right]
\label{eq:BRbqmumu}
\index{Bstomumu@$\text{BR}(B_s\rightarrow \mu^+ \mu^-)$}
\index{Bdtomumu@$\text{BR}(B_d\rightarrow \mu^+ \mu^-)$}
\end{equation}
where 
\beq
X_i=\frac{f_{B_i}^2 }{128 \pi \,m_{B_i} \Gamma_{B_i,H}} \sqrt{1-\frac{4 m_\mu^2}{m_{B_i}^2}}  \; \Longrightarrow \; X_s=5.76\times 10^7\mbox{ and } X_d=3.84\times 10^7,
\eeq \vspace{-5mm}
\bea
\label{eq:bmmFdefapp}
{F}^{(i)}_S =&\;  {m_{B_i}^3\over m_b+m_i} ({ C}_S^{ LL}+{ C}_S^{ LR}-{ C}_S^{ RR}-{ C}_S^{ RL}), \\ 
{F}^{(i)}_P =&\;  {m_{B_i}^3\over m_b+m_i}  (-{ C}_S^{ LL}+{ C}_S^{ LR}-{ C}_S^{ RR}+{ C}_S^{ RL}), \\ 
{F}^{(i)}_A =&\; 2m_{B_i}m_\mu ({ C}_V^{ LL}-{ C}_V^{ LR}+{ C}_V^{ RR}-{ C}_V^{ RL}).
\eea
The width of the heavier $B$ meson is used in the expression \cite{DeBruyn:2012wk} $\Gamma_{B_d,H}\approx\Gamma_{B_d}\equiv \tau_{B_d}^{-1} =4.33\times 10^{-13}$ GeV and $\Gamma_{B_s,H}\equiv \tau_{B_s,H}^{-1}=4.104\times 10^{-13}$ GeV. $m_b$ in these expressions is $m_b(m_b)$ the $\bar{\mbox{DR}}$ renormalization scheme \cite{Dedes:2008iw}.  The values of other parameters can be taken from table \ref{tab:MesonProperties}. The standard model contribution to $B_i\to \mu^+\mu^-$ has been evaluated to QCD NNLO order \cite{Hermann:2013kca} and EW NLO order \cite{Bobeth:2013tba} to be \cite{Bobeth:2013uxa}:
\beq
F^{(i)}_{A,SM} = 0.4690\, \times \frac{4 \alpha_2 V_{tb}^*V_{ti}}{\pi}  \frac{m_\mu m_{B_i}}{v^2}
 ; \; F^{(i)}_{S,SM}=F^{(i)}_{P,SM}=0
\eeq
 which translates to $F^{(d)}_{A,SM}=(1.5-0.6 i)\times 10^{-9}$ and $F^{(s)}_{A,SM}=(-7.8-0.1 i)\times 10^{-9}$.  The form factors in (\ref{eq:bmmFdefapp}) undergo no QCD running \cite{Dedes:2008iw}.

 The observables BR$(B_s\to \mu^+\mu^-)$ and BR$(B_d\to \mu^+\mu^-)$  are evaluated with the \FF\ functions \code{Bstomumu} and \code{Bdtomumu}. We note that \FF's evaluation and compilation of the Wilson coefficients for $B_s\to \mu^+\mu^-$ and $B_d\to \mu^+\mu^-$ is one of the largest bottlenecks in the program.  For more detailed discussions of program speed, see sections \ref{sec:LIL} -- \ref{sec:building}.
\codebox{Bstomumu[$\mu$,Heff]}{evaluate BR$(B_s\rightarrow\mu^+\mu^-)$ for Wilson coefficients at $\mu$\index{Bstomumu}}{0.68} \vspace{-5.8mm}
\codebox{Bdtomumu[$\mu$,Heff]}{evaluate BR$(B_d\rightarrow\mu^+\mu^-)$ for Wilson coefficients at $\mu$\index{Bdtomumu}}{0.68}

\subsection{Neutron EDM}
\label{sec:nEDM}

The electric dipole moment of the neutron can be expressed in terms of the dimension 5 operators (\ref{eq:OpA})--(\ref{eq:OpG}) with $f_1=f_2=d$ or $f_1=f_2=u$. The expression for the neutron EDM (at 1-loop from quarks) can be written as \cite{Hisano:2012sc}, 
\beq
d_n = 0.12^{+0.09}_{-0.06} \; e \left[ (4 d^e_d - d^e_u) + 1.5 (2 d^c_d + d^c_u) \right],
\label{eq:nEDM}
\index{Neutron EDM}
\eeq
where $e\sim0.30$ is the electromagnetic coupling, and the electromagnetic $d^e_q$ and chromomagnetic $d^c_q$ portions can be written in terms of Wilson coefficients as,
\bea
d^e_q =&\; 2\Im\! \left[C_{A,q}^R-C_{A,q}^L \right] \\
d^c_q =&\; 2\Im\! \left[C_{G,q}^R-C_{G,q}^L \right],
\eea
 where the factor of two is correcting for a relative normalization between \FF\ operators and the operators in reference \cite{Hisano:2012sc}.

The RG evolution of the operators relevant for the neutron EDM \cite{Buras:1998raa,Degrassi:2005zd} is performed in \FF\ with the function \code{EDMRG}.  As the Wilson operators (\ref{eq:OpA})--(\ref{eq:OpG}) of interest in \FF\ and \cite{Hisano:2012sc} do not explicitly  contain the quark mass as those in \cite{Buras:1998raa,Degrassi:2005zd} do,  the RG evolved operators must further be scaled by the factor $\lp\frac{\alpha_s(\mu_H)}{\alpha_s(\mu_L)}\rp^{4/\beta_0}$ to account for the running of $m_q$, where the QCD beta function is $\beta_0=11-2 n_F/3$ for $n_F$ light quark flavors.  

\codebox{EDMRG[$\mu_L$,$\mu_H$,$d$list,$u$list]}{RG evolve the EDM Wilson coefficients from the high scale, $\mu_H$, to the low scale, $\mu_L$, with the output in the form $\{d$\code{list},$u$\code{list}$\}$ where for both input and output \code{$q$list}$=\{C_{A,q}^{R},C_{G,q}^{R},C_{A,q}^{L},C_{G,q}^{L}\}$}{0.60}

We stress that the uncertainty on the coefficient in (\ref{eq:nEDM}) is $\order{1}$.  Moreover, while this coefficient's value was determined from QCD sum rules, results in $\chi$PT \cite{Mereghetti:2010kp,Guo:2012vf}, on the lattice \cite{Shintani:2005xg,Berruto:2005hg,Shintani:2006xr,Shintani:2008nt,Alexandrou:2015spa,Shintani:2015vsx}, and even other QCD sum rules evaluations \cite{Pospelov:2000bw}, can vary drastically, so that the uncertainty on that uncertainty is also $\order{1}$ or more.  So any theoretical prediction of the neutron EDM, from \FF\ or otherwise, should be viewed as an order of magnitude estimate at best. 

Further limitations of the \FF\ prediction include: we neglect potentially important contributions from the Weinberg operator \cite{Weinberg:1989dx}, $\CO_W = g_3 f^{abc} \epsilon^{\mu\nu\rho\sigma} G^a_{\mu\lambda}G^{b,\lambda}_{\nu} G^c_{\rho\sigma}$;  and \FF\ is limited to one-loop contributions to the neutron EDM, while it is well-known that sometimes two-loop contributions can dominate \cite{Hisano:2006mj}. 

\codebox{neutronEDM[$\mu$,Heff]}{estimate neutron EDM (in units of  e$\cdot$cm) for Wilson coefficients at $\mu$\index{neutronEDM}}{0.66}

\subsection{$\ell_i\to \ell_j \gamma$}
\label{sec:litoljgamma}

The leading contribution to radiative lepton decays come from the dimension five photon operators, $\CO_A^M$ (\ref{eq:OpA}), with $f_1=\ell_i$ and $f_2=\ell_j$. Although the existence of neutrino oscillations implies that these are nonzero, all standard model lepton flavor violation is proportional to the very small neutrino masses.  In particular, for the radiative decays, a suppression of at least $m_\nu^2/m_W^2< 10^{-23}$ is expected to accompany the decay rate, resulting in total rates that are immeasurably low compared to the reach of foreseeable experiments \cite{Crivellin:2013hpa}.

   Although, more sophisticated treatments are certainly possible, see for instance \cite{Davidson:2016edt}, the effects beyond leading order are typically very small, especially given the current absence of an observed signal.  These radiative decay branching ratios can be simply expressed as \cite{Kuno:1996kv}, 
  \beq
\text{BR}(\ell_i\to\ell_j\gamma) =  \frac{\alpha_{EM} m_{\ell_i}^3}{\Gamma_{\ell_i}} \lp 1 -\frac{m_{\ell_j}^2}{m_{\ell_i}^2} \rp^3 \lp \abs{C^L_A}^2 +\abs{C^R_A}^2\rp 
\index{mutoegamma@$\text{BR}(\mu\rightarrow e\gamma)$}
\index{tautoegamma@$\text{BR}(\tau\rightarrow e\gamma)$}
\index{tautomugamma\@$\text{BR}(\tau\rightarrow \mu\gamma)$}
  \eeq
 In \FF, no standard model contribution is provided, and the very small phase space factor is neglected.   The Wilson coefficients are assumed not to run from the high scale, which is a good approximation.

\codebox{mutoegamma[$\mu$,Heff]}{evaluate BR$(\mu\rightarrow e\gamma)$ for Wilson coefficients at $\mu$\index{mutoegamma}}{0.68} \vspace{-6.8mm}
\codebox{tautoegamma[$\mu$,Heff]}{evaluate BR$(\tau\rightarrow e\gamma)$ for Wilson coefficients at $\mu$\index{tautoegamma}}{0.68}\vspace{-6.8mm}
\codebox{tautomugamma[$\mu$,Heff]}{evaluate BR$(\tau\rightarrow \mu\gamma)$ for Wilson coefficients at $\mu$\index{tautomugamma}}{0.68}

\section{Numerical Wilson Coefficients:  FFWilson}
\label{sec:wilson}

In this section we will describe the contents of the \FFW\ package. One of the main components of \FFW\ is the code to compile the analytic amplitudes. This was discussed already in section \ref{sec:compile}.  Here we will focus on the other components of the \FFW\ package. These include libraries for loop integral evaluation, basic functions to build the amplitudes for faster compilation, and functions to extract the numerical Wilson coefficients at the SUSY scale.

\subsection{Loop Integral Libraries}
\label{sec:LIL}
\index{Loop Integrals}

Analytic expressions for Passarino-Veltman integrals can have ``artificial instabilities'' due to machine precision  when mass parameters are nearly degenerate.  In short, terms  of the form
\beq
 \frac{M_1^2 \ln M_2^2-M_2^2 \ln M_1^2}{M_1^2 - M_2^2} 
 \label{eq:accidentaldegeneracy}
\eeq
 often appear.    While (\ref{eq:accidentaldegeneracy}) is perfectly well behaved when $M_1 -M_2 \to 0$, the machine precision rounding will cause this expression to artificially inflate.   For this reason, \FF\ has two separate running modes: ``Fast'' and ``Acc''.   The difference between these two modes is their treatment of the Passarino-Veltman loop integrals.   All code discussed in this subsection is contained in the  \file{Core/LoopIntegrals} subdirectory.

\subsubsection{``Fast'' Loop Integrals}
\index{``Fast'' Mode}

 When running in \fmd, \FF\ explicitly substitutes exact algebraic degeneracies with their analytic limit through use of overloaded integral functions.  But ``Fast'' mode does not solve the issue of accidental or near degeneracies. The file \file{LoopIntegrals.nb} contains code to compute the limits and generate \file{LoopIntegrals.m}. 
 
 Compiling in \fmd\ typically takes $\order{5\mbox{ min}}$, and less than half that using the precompiled built amplitudes (see section \ref{sec:building}).   Once compiled, evaluating all processes included with \FF\ at a single point in \fmd\ generally takes less than 0.1 seconds.  

\subsubsection{``Acc'' Loop Integrals}
\index{``Acc'' Mode}

In \amd\ (accurate mode),  mass parameters in the loop are checked case-by-case to assess whether there are any near degeneracies or not.   In the case of a near degeneracy, a previously derived analytic expression that uses a Taylor expansion to fourth order in the mass splitting(s) is evaluated.  The file \file{LoopIntegralsAcc.nb} contains code to compute the Taylor expansions and generate \file{LoopIntegralsAcc.m}.    Importantly, \amd\ has a somewhat simplified table of possibilities presented in \file{LoopIntegralsAcc.nb} based on the different loop functions that actually appear in the current array of observables.  This simplification is introduced in order to reduce the number of cases that need to be checked before evaluating, which slightly improves the overall speed.  The integral expressions within \file{LoopIntegralsAcc.m} are actually automatically compiled by the \FFW\ package.
 
 Compiling observables in \amd\ typically takes $\order{3\mbox{ min}}$, and less than half that using the precompiled built amplitudes (see section \ref{sec:building}).  Due to the very large number of loop integrals that need to be evaluated on-the-fly for each observable at every point, \amd\ is roughly a factor of 20 slower than \fmd, and typically takes about 2 seconds to evaluate each point.

\subsection{Building for Faster Compiling}
\label{sec:building}

 Building amplitudes is a way to decrease the compiling time.   When building amplitudes, many of the time-consuming substitutions performed prior to compiling are enacted in advance.  These precompiled amplitudes are stored in a separate directory.  These files are much, much larger than the amplitudes generated by \CA\ as all summations are expanded, many standard model parameters are evaluated, and, in the case of \fmd, long form integral expressions are presented explicitly.
 
 Building amplitudes takes time, while \amd\ is a bit faster, building all amplitudes for both running modes takes on the order of 10 minutes.  However, once this step has been performed once it does not need to be performed again unless:
\begin{enumerate}[label={\alph*)}]
\item standard model parameters (see Sec.~\ref{sec:ffpackage}), such as the CKM matrix, are changed
\item  a new observable is added (although one may build only that observable)
\item  a new model is introduced (amplitudes for other models do not need to be rebuilt)
\item  built amplitudes are removed to save disk space
\end{enumerate}
 For both running modes, building the amplitudes reduces compiling time by more than a factor of two, and has no affect on evaluation time.  These built amplitudes, which are stored in the \file{\{model\}/ObservableAmps/SubAmps} directory, occupy a fair bit of disk space, which scales with additional observables.  The current array of observables  occupies about 0.5 GB  of disk space.  

\codebox{BuildFF[proclist]}{Build all amplitudes in \code{proclist} for both running modes\index{BuildFF}}{0.70}\vspace{-6.9mm}
\codebox{BuildFF[proclist,mode]}{as above, but for running mode \code{mode}}{0.6}\vspace{-6.9mm}
\codebox{CleanBuiltFFFiles[]}{Removes all built amplitudes\index{CleanBuiltFFFiles}}{0.5} \vspace{6.9mm}

Built amplitudes can be removed with \code{CleanBuiltFFFiles[]}.  The command opens a message box requiring the user to type ``Y'' in order to remove the files.  Alternatively, the files can be deleted from the \file{\{model\}/ObservableAmps/SubAmps} directory by hand.  All files in that directory must be removed as compiling will default to attempting to compile the built files if any files exist.

\subsection{Extracting Wilson Coefficients}
\label{sec:FFWilson}

From the amplitudes, \FF\ can extract the numerical Wilson coefficients for a model.    The Wilson coefficient extraction is handled by the function \code{FFWilson}, which loops over all compiled processes and evaluates the amplitudes numerically.  
 \codebox{FFWilson[VariableList]}{extract nested table of processes and numerical Wilson coefficients from the analytic amplitudes\index{FFWilson}}{0.65}
 where \code{VariableList}  is the output from \code{CalcSpec} or \code{FFRunFile} (see \ref{sec:CalcSpec}) defined by the model specific global variable \code{\$FFCompileVarList} within \file{\{model\}/CompileAmps.m}.  The \code{VariableList} must be given as real or complex numbers as appropriate for the variable in question, if this is not done, then the compiled amplitudes will not be used and the code will run extremely slowly (potentially with errors).
  
 The output of \code{FFWilson} is a nested list of the form:
 \blockc{\{\{\{process name, external state, topology type\}, \{scale, $\sum_i C_i \CO_i$\}\},...\}} 
 where process name (an element of \code{\$FFProcessList}), the external state are the fields, e.g., $\{\{``s",``d"\},\{``s",``d"\}\}$, the topology type is either ``boxes'', ``penguins'', or ``wavefncorr'' (wave function corrections), and scale is the scale at which the Wilson coefficients are evaluated (this is typically the $\order{\mbox{TeV}}$ SUSY scale).  The sum over operators includes all those relevant for the process, in the basis (\ref{eq:OpA})--(\ref{eq:OpT}), with $C_i$ being the numerical Wilson coefficients.

\section{The Flavor-Violating MSSM:  FFModel}
\label{sec:MSSM}

\FFM\ is the name given to the model-dependent part of the code that reads in spectra, compiles Wilson coefficients, and performs other miscellaneous model-dependent tasks.  Currently, the only \FFM\ provided with \FF\ is the flavor-violating MSSM, i.e.\ \FFM$=$\code{MSSM}. It resides in the subdirectory  \file{MSSM/}. In principle, \FF\ allows for additional  \FA\ models to be used with the \CA\ package to produce the amplitudes.  The onus is then on the user to link the new model with the core code of \FF, by constructing new \FFM\ code.  In principle, only the spectrum calculator, some form of I/O for obtaining the parameters, dictionary of standard model field names for \CA, and model specific compiling definitions are needed.  Although \FF\ defaults to the MSSM, a new model can be used by setting \code{FormFlavor`\$FFModel=\{Model Directory\}}\index{FormFlavor`\$FFModel}\index{\$FFModel} prior to loading \FF.  In this section, we describe the spectrum calculator and SLHA2 I/O, the RG created with the aid of SARAH, and custom functions that come with the MSSM \FFM.

\subsection{Spectrum Calculator \& SLHA2 I/O}
\label{sec:CalcSpec}

The MSSM-specific part of \FF\  allows for three different input modes:
\begin{enumerate}
\item \FF\ can read raw {\it input} SLHA2 files  (basically, MSSM soft masses and Higgs parameters) and perform a basic tree-level calculation of the spectrum and mixing angles. In more detail: an SLHA2 input file contain the soft parameters, notably 3 x 3 mass squared matrices for $\tilde m_Q^2$, $\tilde m_U^2$, $\tilde m_D^2$, $\tilde m_L^2$, and $\tilde m_E^2$ in the super CKM basis.  For input, these require the following SLHA2 blocks: \code{MSQ2}, \code{MSU2}, \code{MSD2}, \code{MSL2}, \code{MSE2}, \code{TU}, \code{TD}, \code{TE}, \code{MSOFT} (parameters $\{1,2,3,21,22,25\}=\{M_1, M_2,M_3,M_{H_d}^2,M_{H_u}^2,\tan\beta\}$, \code{HMIX} ($1=\mu$), \code{ALPHA} ($1=\alpha$ the CP even Higgs mixing angle), \code{MASS} ($\{25,36\}=\{m_h,m_A\}$ only) are required.  The blocks  \code{IMMSQ2}, \code{IMTU}, etc., are also used for the sfermion mass matrices and $A$ terms, however, if not provided these parameters are set to zero with no error.  See \cite{Allanach:2008qq} for detailed descriptions for the conventions of these parameters.  We will note that in SLHA2 conventions the matrix $\tilde m_Q^2$ is aligned with the down-type quark basis, i.e., the 1, 2, 3 rows and columns point in the $\{d,s,b\}$ direction (as opposed to the $\{u,c,t\}$ directions). If \code{ALPHA} is not provided, the code defaults to the alignment limit, that is $\alpha=\beta-\frac\pi2$.

\item  Alternatively, \FF\ can read {\it output} spectrum files containing masses and mixings in the SLHA2 format, which could include threshold corrections coming from one of the many public MSSM spectrum calculating codes.  An SLHA2 output file contains the physical masses, notably 6 x 6 rotation matrices to connect the mass eigenbasis to the flavor eigenbasis.  The six physical up-type squarks are ordered by ascending mass, i.e., the lightest squark is called $\tilde u_1$ and is identified with PDG ID 1000002, and the first row of the 6 x 6 rotation matrix indicates the flavor composition of this eigenstate (basis $\{\tilde u_L,\tilde c_L,\tilde t_L,\tilde u_R,\tilde c_R,\tilde t_R\}$).  Analogous definitions apply for down-type squarks, sleptons, and the three flavors of sneutrino.  The following SLHA2 blocks are required, \code{MSOFT} (only needs $25=\tan\beta$), \code{MASS} (all super particle masses as well as Higgs masses), \code{HMIX} ($1=\mu$ only), \code{ALPHA} ($1=\alpha$), the 4 x 4 neutralino mixing matrix \code{NMIX}, the two 2 x 2 charging mixing matrices \code{UMIX} and \code{VMIX}, the 6 x 6 sfermion mixing matrices \code{USQMIX}, \code{DSQMIX}, \code{SELMIX}, as well as the 3 x 3 sneutrino mixing matrix \code{SNUMIX}.  \FF\ additionally requires the $A$-terms, as these are couplings, so \code{TU}, \code{TD}, and \code{TE} are required.  Again, \code{IMUSQMIX}, \code{IMTU}, etc., are used, but are simply set to zero if not provided.

\item Finally, \FF\ has an internal soft mass and Higgs parameter format that it uses via the routine \code{CalcSpec} (see below), and this can also be input directly to the program:
\begin{equation}\label{eq:csformat}
\begin{minipage}[c]{0.75\textwidth} gaugino[$\{M_1,M_2,M_3\}$],higgs[$\{\mu,M_h,M_A,\tan\beta,\alpha\}$],squarkQLL[$m_Q$], squarkURR[$m_U$],squarkDRR[$m_D$],sleptonLL[$m_L$],sleptonRR[$m_E$], Au[$A_u$],Ad[$A_d$],Ae[$A_e$] \end{minipage}
\index{CalcSpec!Input Format}
\end{equation}
 This mode is useful for quickly generating grids without needing to create SLHA2 files.
\end{enumerate}

 The primary function for reading SLHA2 files is \code{FFReadFile}.  
 \codebox{FFReadFile[file]}{read in parameters from SLHA2 \code{file} and output spectrum in format to be fed to compiled amplitudes}{0.7}
This function automatically decides whether the file is SLHA2 input or output. In the case of SLHA2 input, \code{FFReadFile} automatically calls a simple spectrum calculator, called \code{CalcSpec}, 
  \codebox{CalcSpec[CS]}{convert soft parameters in super CKM basis (input form shown in (\ref{eq:csformat})) to output spectrum to be fed into compiled amplitudes\index{CalcSpec}}{0.78} 
  that will convert the SLHA2 input to a spectrum that can be directly fed into the compiled amplitudes.   The function joins the soft parameters with the $D$-term contributions,  manipulates the $A$-terms and $\mu$-parameter expressions to compute the 6 x 6 mixing matrices and eigenvalues. Importantly, the spectrum calculator sets all neutralino eigenvalues to be positive and uses a complex mixing matrix.  No matter the read in format, the output of  \code{FFReadFile} and \code{CalcSpec} is put into the format of \code{\$FFCompileVarList}\index{\$FFCompileVarList} defined within \file{\{model\}/CompileAmps.m}.\footnote{In truth, the output needs to be flattened to match \code{\$FFCompileVarList}, but the output is more readable prior to flattening.  This flattening is automatically performed by \FFW.}
   
Once these parameters are evaluated, the resulting spectrum may be output to an SLHA2 file using the \code{FFWriteFile} command.
  \codebox{FFWriteFile[CalcSpecOutput,file]}{ takes the \code{CalcSpec} output and writes to \code{file} in SLHA2 format\index{FFWriteFile}}{0.5}
   
  Both the spectrum calculator and the SLHA2 input and output routines are contained in \file{MSSM/CalcSpec.m}.  \FF\ comes with equivalent SLHA2 input and output files, called \code{ExampleSLHA2in.dat} and \code{ExampleSLHA2out.dat}, that are parameter points in a $Q$-class model of extended gauge mediation \cite{Evans:2013kxa,Evans:2015swa}.
  
\subsection{RG Evolution}
\label{sec:RGE}

 \FF\ additionally contains a simple function for performing renormalization group evolution.  This evolution contains full 3 x 3 running at one loop level including CP violating phases and applies BMPZ QCD threshold corrections \cite{Pierce:1996zz}.  The basic RG equations were derived using the SARAH package \cite{Staub:2010jh}.  As the RG was designed with  flavor physics in mind,  many features for a more precise spectra determination, such as two-loop Higgs corrections, back-and-forth running for better specification of scale, and threshold corrections (except for the QCD BMPZ threshold corrections), are not included in this module.  
  \codebox{RGFile[infile,outfile,$\mu$]}{RG evolve high scale \code{infile} down to scale \code{$\mu$} and output \code{outfile}\index{RGFile}}{0.5} 
 
  The RGE is contained in the file \file{MSSM/RGE.m}, although some of the SLHA2 I/O is defined in  \file{MSSM/CalcSpec.m}.  Importantly, the RGE is not used by any other portion of the code.  A user creating a new \FFM\ would not need to create an RGE for that model.

\subsection{Custom User Functions}
\label{sec:CUF}

There are two simple places where one can define user functions to be accessed by the code.  The first is the file \file{Core/UserCore.m}, which contains no functions in the default release of \FF.  The second place that one can define new user functions is \file{\{model\}/UserModel.m}.  Within the \file{MSSM/UserModel.m} file, several functions have been defined that are useful  chains of code.  In addition to the \code{FFfromSLHA2} and \code{FFConstraintsfromSLHA2} defined in section \ref{sec:FFeval}, there is \code{FFWilsonfromSLHA2}, which uses the \code{FFWilson} command of section \ref{sec:FFWilson}.
\codebox{FFWilsonfromSLHA2[file]}{load SLHA2 \code{file} and run \code{FFWilson} using current active running mode}{0.58}
 \vspace{-4.1mm}
\codebox{FFWilsonfromSLHA2[file,mode]}{ as above, but for  running mode \code{mode}\index{FFWilsonfromSLHA2}}{0.60}

\section{The CalcAmps Package}
\label{sec:CalcAmpsP}

 One of the most important and distinguishing features of \FF\ is the ability to calculate one-loop Wilson coefficients for new flavor and CP observables and/or new models, entirely from scratch. These tasks are performed by the \CA\ package, based on the machinery of \FA\ \cite{Hahn:2000kx} and \FC\ \cite{Hahn:1998yk}. The various routines used by \CA\ are described in section \ref{sec:CalcAmps}, while section \ref{sec:newobs} contains a detailed explanation of how  a user can add new observables to the program.  The wrapper package \file{\{model\}/CalcAmps/CalcAmps.m} loads the main code in  \file{Core/CalcAmpsPackage.m} and links to the model-specific definitions in \file{\{model\}/CalcAmps/CalcAmpsModel.m}. 
 \index{CalcAmps}
 
\subsection{CalcAmps}
\label{sec:CalcAmps}
  
  The main routine to generate new processes is also called \code{CalcAmps} and is found in the file \file{Core/CalcAmpsPackage.m}.   
  
     \codebox{CalcAmps[proc,$\CO1$,$\CO2$]}{Calculate amplitudes relevant for process \code{proc} involving external states contained in $\CO1$ and $\CO2$.  This is the main routine of the \CA\ package.\index{CalcAmps}}{0.67}
     
\noindent It takes as input a process name (e.g., ``K-K mixing"), and two lists of external SM fields, which define the operator basis for the observable. The syntax is $\{f_1,f_2\},\{f_3,f_4\}$ for  the set of 4-fermi operators with Dirac indices contracted between the braces $(\bar f_1(\dots)f_2)(\bar f_3(\dots)f_4)$ and $\{f_1,f_2\},\{b_1\}$ for a dipole-type operator $(\bar f_1(\dots)f_2)(b_1)$.  Note that the order of the particles is important.  Additionally,  the $f_i$ should only be SM particles, anti-particles will not be accepted. Currently only 3- and 4-body processes are supported, so these lists must be length 2 and 2, or 2 and 1, or 1 and 2.  Note that $1\to3$ processes, such as $K\to \pi \nu\nu$, are expressed as 2 and 2, i.e., $\{d,s\} \{\nu,\nu\}$.

 One of the most important subroutines called by the \code{CalcAmps} function is the function \code{GenerateDiagrams}. 
   \codebox{GenerateDiagrams[$\CO1$,$\CO2$]}{Generate all diagrams connecting initial and final states $\CO1$ and $\CO2$.\index{GenerateDiagrams}}{0.60}
\noindent  \code{GenerateDiagrams} is a general purpose routine that takes as input the initial and final state particles and then draws all one-loop diagrams connecting them.   These diagrams are further classified into topologies (boxes, penguins and wave-function corrections) that are relevant for 3- and 4-particle processes. (Currently, 5+ particle processes are not supported.)  Default options for \code{GenerateDiagrams} include: \code{Model$\to $"FVMSSM"}, other models are in principle possible; and \code{GenerateSM$\to$False}, this eliminates all diagrams involving only SM fields, since these are accounted for (often with much higher precision than 1-loop) by the observable functions discussed in section \ref{sec:obs}.  In order for \code{GenerateSM$\to$False} to work, the user has to specify all the SM fields with the list, \code{SMlist} defined in \file{\{model\}/CalcAmps/CalcAmpsModel.m}.  \code{GenerateDiagrams} returns a list of diagrams in FeynArts format indexed by their topology type, e.g., 

\vspace{-3mm}\blockc{$\{\{$``boxes",box diagrams$\},\{$``penguins",penguin diags$\},\{$``wavefncorr'', wfc diags$\}\}$}

\code{CalcAmps} uses subroutines defined within the package to turn the ``DiracChains" of \FC\ (basically Gamma matrices sandwiched between external state spinor wavefunctions, e.g., $\bar v(p_1)\gamma^\mu u(p_2)\bar u(p_3)\gamma_\mu v(p_4)$)  into the Wilson operators used throughout \FF. It is careful to include a factor of 2 when the initial and final states are CP conjugate as is the case for $\Delta F=2$ observables, i.e., the operator $\langle \bar d s|(\bar s d) (\bar s d)|d\bar s\rangle = 2(\bar u v)(\bar v u)$.  \code{CalcAmps} will also apply the Gordon identity to extract the magnetic dipole operator from amplitudes of the form $(\bar u v)\epsilon\cdot k$.\index{DiracChain}\index{Gordon Identity}\footnote{In order to have appropriate signs in front of the relevant Wilson operators, \CA\ must address two technical subtleties concerning fermion ordering within \FA\ and \FC. First,  the diagrams generated with \FA\ can result in different signs in \FC\ when the order of the initial and/or final state fermions are changed.  For the operators $(\bar f_1(\dots)f_2)(\bar f_3(\dots)f_4)$, the correct order is $|f_2,\bar f_1\rangle$ and $|f_3,\bar f_4\rangle$ for the initial and final states.  Then $\langle f_3,\bar f_4|C (\bar f_1(\dots)f_2)(\bar f_3(\dots)f_4)|f_2,\bar f_1\rangle  = C\bar v_1 (\dots) u_2 \bar u_3(\dots) v_4$ with no additional signs.  The second technical subtlety is that \FC\ contains an option \code{FermionOrder} (which defaults in \FC\ to \code{Automatic}).   A different choice of \code{FermionOrder} results in a different DiracChain related to one another by Fierz rearrangement identities, which can result in different signs or even different operator bases, potentially leading to an incorrect identification of the Wilson coefficient. \code{CalcAmps} sets \code{FermionOrder=$\{2,1,3,4\}$} for a four-fermi interaction, and \code{FermionOrder=$\{2,1\}$} for $2\to 1$ process.  This hardwired choice is a function of the order of the initial and final states in the call to \code{GenerateDiagrams}, which is called automatically from \code{CalcAmps} to avoid user error.  The user only has to specify the operator basis and in principle these details will be taken care of automatically. \index{FermionOrder}}

Another important feature of \code{CalcAmps} is the third generation dominant approximation. Currently, the option \code{ThirdGenDominance$\to$True} is set by default for \code{CalcAmps}.  In principle, one can set \code{ThirdGenDominance$\to$False}, but this has not been tested and may result in extremely slow evaluation times. In the 3rd gen dominant approximation, all 1st/2nd generation SM fermion masses are set to zero, with the exception of $m_\mu$ which is preserved to keep important terms in $B_q\to\mu\mu$.  This zeroing is enacted to significantly speed up the computation of the loop diagrams, both in \CA\ and \FF.  
\index{ThirdGenDominance}

Formally, terms that are suppressed by powers of $m_f/M_W$ or $m_f/M_{SUSY}$ are actually higher dimension (e.g., dimension 8) operators. To remove these, all the Mandelstam invariants $S$, $T$, etc., are set to zero in the  computation.  Essentially, \FF\ evaluates the Wilson operators at zero external momentum.  In order to further simplify expressions and reduce evaluation time, \code{CalcAmps} also drops all powers of $m_b$ and $m_\mu$ unless they are coming from Yukawa couplings that do not appear suppressed by $\tan\beta$. Finally, there are a number of MSSM-specific simplifications (everything preceding this is in principle model-independent or at least 2HDM model independent, provided one uses the same notation as default FeynArts MSSM for the fermions and their masses) to further speed up the later compilation and numerical evaluation of the Wilson coefficients.  All of these model specific simplifications are contained in the file \file{\{model\}/CalcAmps/CalcAmpsModel.m}

 The output of \code{CalcAmps} should be stored in a file with the use of \code{WriteAmp} which should automatically put the file into the \file{\{model\}/ObservableAmps} directory that contains all amplitudes.    It is from this directory that \FF\ accesses the amplitude files.
 \codebox{WriteAmp[amp,file]}{Write an \code{amp} generated by \CA\ to \code{file} within the \file{\{model\}/ObservableAmps} directory.\index{WriteAmp}}{0.68}

 For ease of use, there is a simple default \CA\ front-end notebook in the \file{MSSM/} directory,  \file{\{model\}/CalcAmps/CalcAllAmps.nb}, which can be used as a template for adding new processes.

\newcommand{\miniblock}[1]{\begin{center} \vspace{-2mm}  \hspace{3mm} \fcolorbox{black!10}{black!10}{\begin{minipage}[c]{0.80\textwidth}
#1\end{minipage} 
}\end{center} }
\newcommand{\miniblockc}[1]{\miniblock{\begin{center} #1\end{center} }}

\subsection{Adding New Observables}
\label{sec:newobs}
 \index{Adding a new observable}

 New observables can be added to \FF\ rather easily.  First, observables that depend on the same Wilson coefficients, for instance $\Delta m_{K}$ \& $\epsilon_K$, can be added within the same observable function, and do not need to be separately produced in \CA.  However, this subsection is meant to act as a tutorial for adding observables that depend on new Wilson coefficients.    By following the instructions provided in this subsection, the user should be able to add their own observables to \FF.
 
 \begin{enumerate}
 \item {\bf Generate amplitude with \CA} -- The \CA\ package should be used to generate an amplitude file.   The file \file{\{model\}/CalcAmps/CalcAllAmps.nb} contains several existing processes that  can be mirrored in order to determine the initial and final state definitions.  Generating this amplitude will typically take a little bit of time, but once generated the amplitude should be stored as a \file{*.m} file in the \file{\{model\}/ObservableAmps} directory with the use of \code{WriteAmp}.  (We will refer to this amplitude file later on as \file{\{X\}.m}.) It is recommended that the user gives an intuitive name to both the process and the amplitude file.   Note: \CA\ and \FF\  are not designed to be run simultaneously.   If one package is loaded, the \mma\ kernel should be quit in order to load the other.
 
 \item {\bf  Create and link an \file{Obs\{X\}.m} file} --  With the amplitude stored within the \file{Core/Observables} directory, the user should create an \file{Obs\{X\}.m} file for the process of interest.  It is recommended that the user copy an existing file that is most similar to the process of interest to use as a template. Within \file{Core/FFObservables.m},  the user should add a line,
 \miniblockc{\code{Get[FormFlavor`\$FFPath<>"/Core/Observables/Obs\{X\}.m"]}}
  within an appropriate subsection (or create a new subsection if no appropriate one exists).   This will ensure that when the user loads \FF, the new observable will also be loaded.
  
  \item {\bf  Modify \file{Obs\{X\}.m} for the process of interest} --  All observable files follow a simple convention to facilitate the addition of new observables.   We will present the individual portions and what needs to be modified in the order they appear within the files.
  
   \begin{enumerate}
\item {\bf Linking} -- The first block contains several lines of code, but only the first and last need to be modified.   The first line is 
\miniblockc{\code{TempAmpFileName=\{X\}.m;}}
The user should set this to the filename used in step 1.  This line and the subsequent code informs \FF\  of the name of the amplitude file within the \file{\{model\}/ObservableAmps} directory, and appends the filename to \code{\$FFAmpFileList}\index{\$FFAmpFileList}.   The next line automatically accesses the amp file to extract the process that was defined in step 1, and  appends the process name to \code{\$FFProcessList}\index{\$FFProcessList}.  The last line
\miniblockc{\code{ObservableFunction[TempProcessName]=\{observable function\};}}   
needs to be modified by the user so that the name of the observable function defined in the final block will be linked to the process.

\item {\bf  Observable details} --  The next block contains details on the current experimental and theoretical status of the observable, as well as its name within \FF.   The first line is \code{TempObsName="NAME";}, and should contain the name as you want it to appear in \FF. To reduce user error, \code{TempObsName} is used in every block, but the last line \code{FF\{X\}Name=TempObsName;} defines a unique identifier for the observable's name that will be called again later.   The definitions of \code{FFObsClass}, \code{FFExpValue}, \code{FFExpUnc}, and \code{FFSMUnc} should be introduced and assigned the appropriate values.   \code{FFSMValue} also appears here, however, its definition is typically not used in the code.  See section \ref{sec:obs} for more details about these definitions.    The second to last line for each observable, \miniblockc{\code{AppendTo[\$FFObsNameList,TempObsName];}}  informs \FF\ of the name and presence of the observable.  Multiple observables originating from the same process file can appear in this block.  The user is encouraged to document these with original references.
   
  \item {\bf  Observable specific pre-evaluations} --  An optional third block contains any code that is specific to the observable, but is pre-evaluated in the interest of expediting the runtime calculation of \FF.  Often, it may be in the user's best interest to separate a detailed calculation that depends on standard model parameters from the bulk of the evaluation so that it will not be reevaluated each time the observable's subroutine is passed new numerical Wilson coefficients.

\item {\bf   Observable function} --   The final block of code within each \file{Obs\{X\}.m} file is the observable function.  This is the all important block of code that translates Wilson coefficients into flavor observables.   The function should be given an intuitive, unique name and referred to in the previous block.  Wilson coefficients can be extracted by \code{Coefficient[wilson,$\CO$];} where $\CO$ is the operator name within \FF, e.g., \code{OpV[$``L",``L"$][$\{``b",``s"\},\{``b",``s"\}$]} (see eqs.~\ref{eq:OpA}--\ref{eq:OpT}).   The final line of this function is a list of observable names and values, e.g., \miniblockc{\code{\{\{FFObs1Name,Obs1val\},\{FFObs2Name,Obs2val\},...\}\}}}
 where \code{FFObs1Name} is the name of the observable defined in the second block, and \code{Obs1val} is the numerical value of the observable as determined by the observable functions code.  Any new SM parameters should be added to \file{Core/SMParameters.m}.   Again, the user is strongly encouraged to provide detailed references for the origin of all expressions within the observables code.

    \end{enumerate}
 \end{enumerate}

\section*{Acknowledgements}
\noindent
We are very grateful to W.~Altmannshofer, A.~El-Khadra, Z.~Ligeti, and J.~Zupan for invaluable discussions.  
We especially thank M.~Misiak for providing code to make \FF's evaluation of $b\to s\gamma$ possible.  
We are also grateful to A.~Thalapillil for collaboration in the early stages of this work. 
The work of DS is supported by DE-SC0013678.

\appendix

\newcommand{\chk}{{\color{green}\surd\color{black}}}
\newcommand{\chko}{{\color{orange}\surd\color{black}}}
\newcommand{\xx}{{\color{red}\code{X}\color{black}}}

\section{Comparing FormFlavor to Other Public Codes}
\label{app:comp}

\begin{table}
\begin{center}
\begin{tabular}{l|cccc}
    Evaluation            & \FF & \SF & \FK & \SP \\
                  \hline
Automated One-Loop & $\chk$ & $\xx$ & $\chk$ & $\xx$ \\
Chiral Resummations & $\xx$ & $\chk/\xx$ & $\xx$ & $\xx$ \\
Double Higgs Penguins & $\xx$ & $\xx^*$ & $\chk$ & $\chk$ \\
Wilson Coefficient Running& $\chk/\xx$ & $\chk$ & $\chko$ & $\xx$\vspace{5mm} \\

Inputs  & \FF & \SF & \FK & \SP \\ \hline
 Non-MFV General MSSM & $\chk$ & $\chk$ & $\chk$ & $\chk$ \\
Soft Parameter Input (SLHA2 in)& $\chk$ & $\chk$ & $\chk$ & $\chk$ \\
Full Spectra Input  (SLHA2 out)& $\chk$ & $\xx^\dagger$ & $\xx$ & $\xx$ \\
 Messenger/GUT Scale Input & $\chk$ & $\xx$ & $\chk$ & $\chk$ \\
Threshold Corrections & $\chko/\xx$ & $\xx$ & $\chk/\xx$ & $\chk$ 

\end{tabular}
\end{center}
\caption{\label{tab:codecomp}  A $\chk$ means the feature is included, $\chko$ means the feature is partially included, while an $\xx$ means that it is not.  $\chk/\xx$ means the feature is included and can be enabled/disabled.  \newline \newline \footnotesize
\qq$^*$Although \SF\ does not explicitly include double Higgs penguins, these are presumably accounted for at some level when chiral resummation is enabled. \newline
\qq$^\dagger$The \SF\ 2.5 manual states that it also accepts ``output files produced by other public libraries calculating various aspects of the MSSM phenomenology," but it is not clear what this means from reading the rest of the \SF\ manual and reading the \SF\ code. It appears to not use the threshold corrected masses and mixings, but still performs the tree-level diagonalization itself.
}
\end{table}

There are several existing public codes that calculate flavor and CP observables in the MSSM. In this brief appendix, we will present a validation of FormFlavor against some of these codes.  Our focus will be on the few other codes that can handle a general MSSM spectra without mandating minimal flavor violation, and that compute a broad range of flavor and CP violating observables. These are: \SP\ \cite{Porod:2011nf}, $\mathtt{SARAH-SPheno-FlavorKit}$ (\FK\ for short) \cite{Porod:2014xia}, and \SF\ \cite{Crivellin:2012jv}. Other codes include: \code{SuperIso} \cite{Mahmoudi:2008tp}, \code{NMHDECAY} \cite{Ellwanger:2004xm}, \code{MicrOMEGAs} \cite{Belanger:2013oya}, \code{SusyBSG} \cite{Degrassi:2007kj}, \code{SuperLFV} \cite{Murakami:2013rca}, \code{SuSeFLAV} \cite{Chowdhury:2011zr}, \code{ISAJET} with \code{ISATOOLS} \cite{Paige:2003mg}, and \code{flavio} \cite{Straub:flavio}.  

Table \ref{tab:codecomp} compares various features of these four codes. For instance, whether the code can automatically compute the one-loop Wilson coefficients from scratch, or whether these loop functions are hard-coded.  Some codes include effects beyond one-loop order such as chiral ($\tan\beta$) resummations \cite{Crivellin:2011jt} or double Higgs penguins \cite{Buras:2001mb}.  Other features include the QCD RG evolution of SUSY scale Wilson coefficients down to the low scale of the flavor or CP violating observable.  In some cases, this evolution can affect the ultimate result by as much as a factor of three or more. As far as we can tell, \SP\ and \FK\ do not run the Wilson coefficients with the QCD beta functions from the SUSY scale to the low scale (passing through the top threshold).  For some observables,  but not all (e.g., not for kaon mixing), \FK\ includes running from $m_t$ (e.g., input is assumed to be at the scale $m_t$) to the low scale.  \SP\ says in its manual that all Wilson coefficients are evaluated with couplings at the scale $m_t$. No mention is made of running down to the low scale.  \SF\ and \FF\ both have running from the SUSY scale (generally defined to be some average of the gluino and squark masses) to the low scale.  Finally, there are various types of inputs that the codes could possibly accept: messenger-scale soft parameters, weak-scale soft parameters, masses and mixings with or without threshold corrections.  Features of the four codes are summarized in table \ref{tab:codecomp}.

\subsection{Comparison of the codes}

 We will now describe a more detailed, quantitative comparison of \FF\ against other public codes, specializing further to \FK\ v2.53 and \SF\ v4.8.6, because all three of these have the option of taking soft parameters as inputs and computing tree-level masses and mixings. This allows our comparison of the Wilson coefficients, QCD RG and flavor observable functions to be isolated from the complicated issue of threshold correcting the SUSY spectrum. \SP\ does not have the capability of turning off threshold corrections, so, although examined, we decided not to include it in the comparison.  We will note that a lot of disagreement was found between \SP\ and  the other codes; however, it is difficult to disentangle these from the threshold corrections, and especially regions of parameter space where the threshold corrections would not converge.
 
In \SF,  chiral resummations ($\tan\beta$ resummation) can be toggled on and off to different orders of resummation. Since \FF\ and \FK\ do not have this feature currently, we will compare against \SF\ with chiral resummations turned off.  The different programs have (slightly) different input parameter choices for all of the flavor observables. 

The parameter space of the general flavor and CP violating MSSM is enormous. To compare the codes in a simple and presentable fashion, we choose the following two lines through the parameter space:

\begin{itemize} 
\item Along the ``LLRR line",  the $A$-terms are zero and we turn on equal deformations to all the sfermion soft mass-squareds 
\beq
m^2_{QLL}=m^2_{URR}=m^2_{DRR}=m^2_{LLL}=m^2_{ERR}=m^2{\bf 1}_3 + \left(\begin{matrix} 0 & \delta m^2 &  \delta m^2 \\ ( \delta m^2)^*\!\! & 0 &  \delta m^2\\ ( \delta m^2)^*\!\! &  ( \delta m^2)^*\!\! & 0\end{matrix}\right)
\eeq

\item Along the ``LR line", the soft mass-squareds are diagonal and we are turning on
\beq
A_u=A_d=A_\ell= \left(\begin{matrix} 0 & A & A \\ A & 0 & A \\ A &  A & 0\end{matrix}\right)
\eeq

\end{itemize}

\noindent  We have set all the gaugino soft masses and $\mu$ to 1~TeV, $m=500$~GeV, $m_A=2$~TeV, $\tan\beta=10$ and $\arg A=\arg\delta m^2=0.1$. 

\begin{figure}
\centering
\includegraphics[scale=.65]{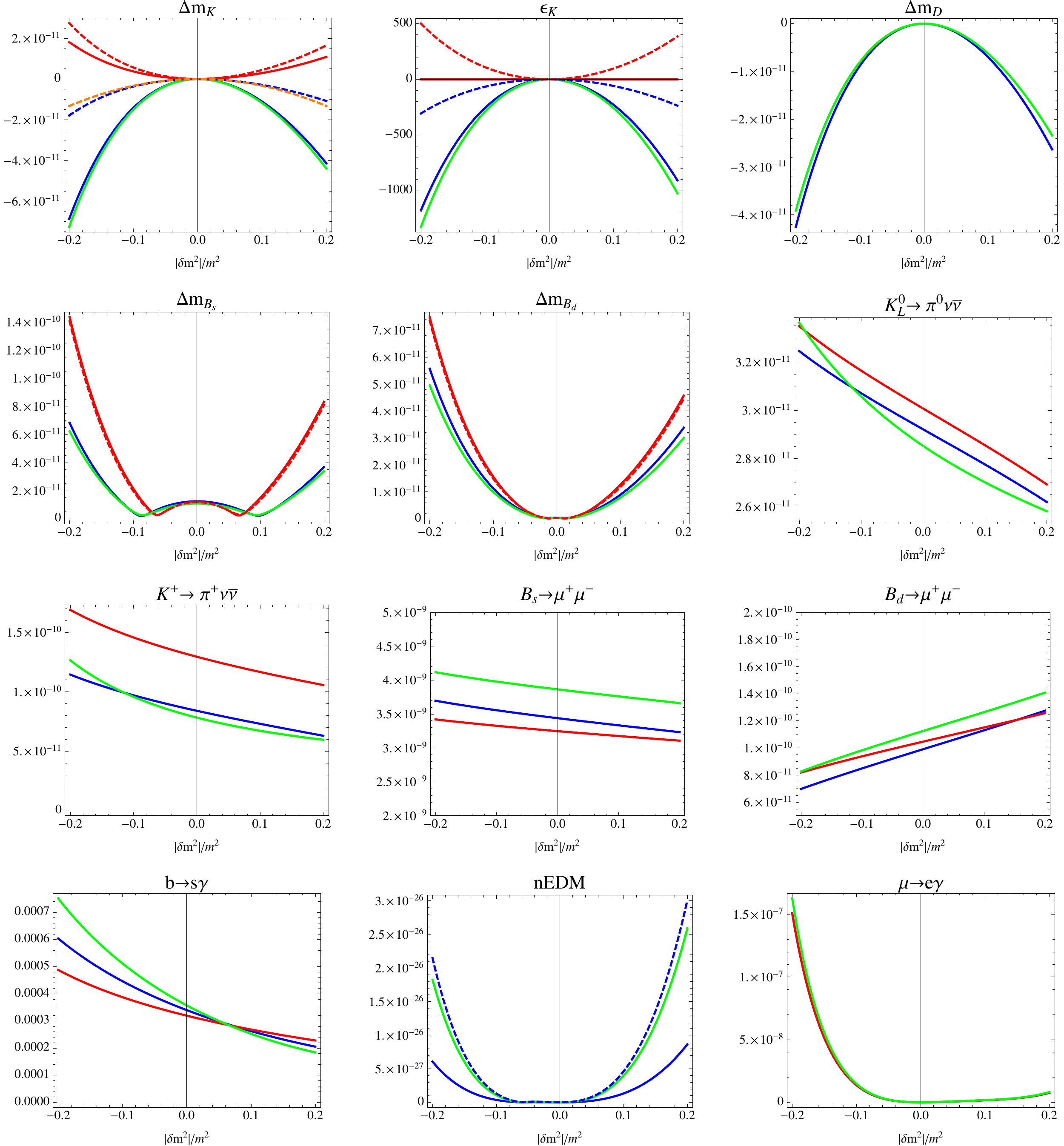}
\caption{\small{ LLRR line as described in the text. Blue is \FF, green is SUSY\_Flavor, and red is FlavorKit. Solid lines indicate the flavor observable computed with the respective code ``out of the box" i.e.\ with no modifications to the default settings. Dashed lines, where present indicate non-default options or modified source code as described in the text. }}
\label{fig:finalLLRR}
\end{figure}

\begin{figure}
\centering
\includegraphics[scale=.65]{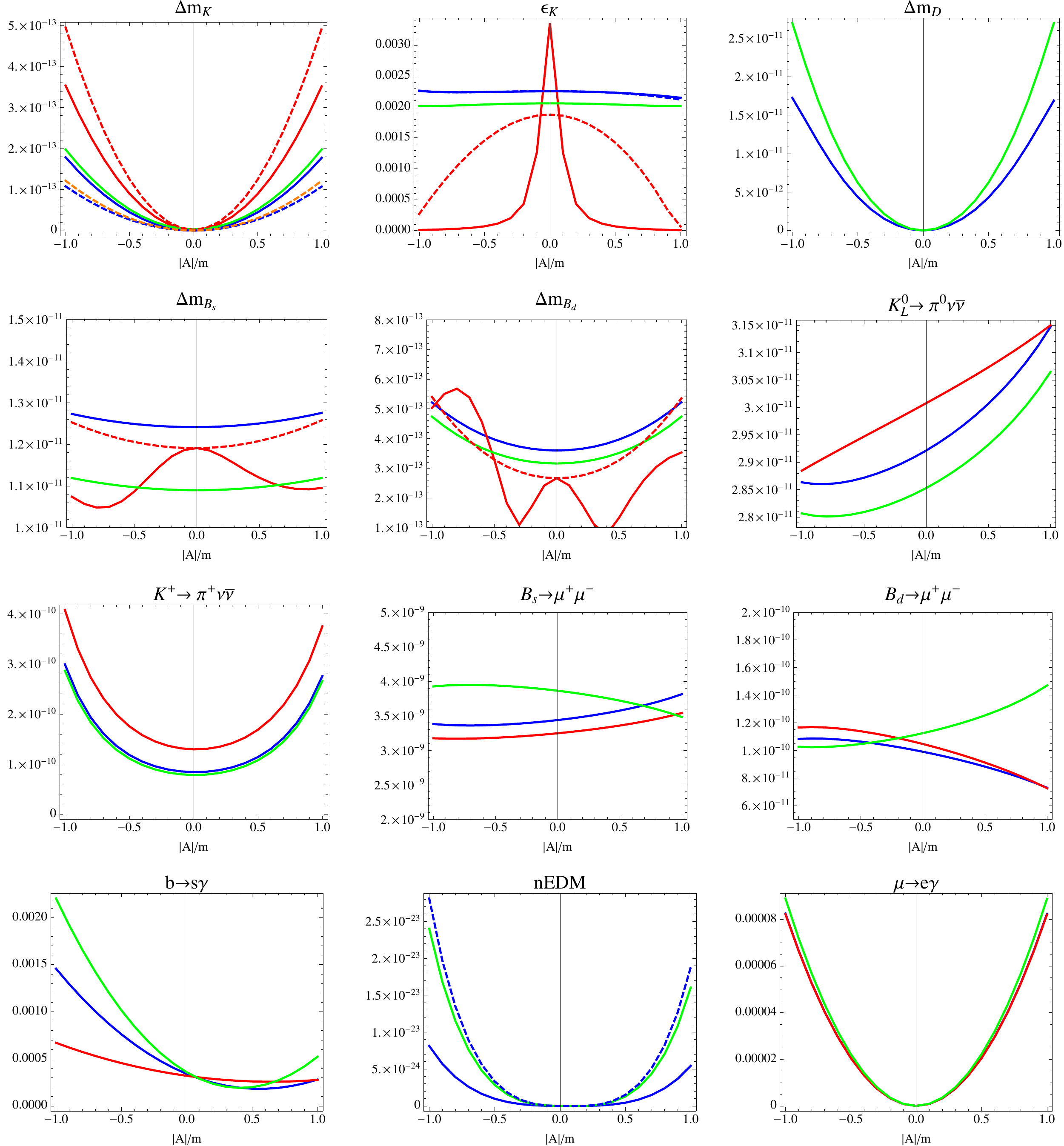}
\caption{\small{ LR line as described in the text. Blue is \FF, green is SUSY\_Flavor, and red is FlavorKit. Solid lines indicate the flavor observable computed with the respective code ``out of the box" i.e.\ with no modifications to the default settings. Dashed lines, where present indicate non-default options or modified source code as described in the text.}}
\label{fig:finalLR}
\end{figure}

Shown in figs.~\ref{fig:finalLLRR} and \ref{fig:finalLR} are the comparison of the three codes along the LLRR line and the LR line respectively. The solid lines indicate the respective code run ``out of the box," i.e., with all default input parameters (except for disabling chiral resummations in the case of \SF) and no other modifications to the codes. We see that all three codes generally agree well on the $\Delta F=0$, 1 observables; while for $\Delta F=2$ observables, \SF\ and \FF\ agree well but \FK\ disagrees strongly with both.

In some instances (\SF's neutron EDM and \FK's $\Delta m_K$ and $\epsilon_K$), the QCD RG from the SUSY scale to the low scale is not included. In these cases, we display a blue dashed line where the QCD RG in \FF\ has been turned off in order to better compare \FF's  evaluation with those of the other code.  The ability to  disable the QCD RG for comparison and debugging purposes is a useful option only found in \FF. 

Inspecting all three codes, we find that the flavor observable functions are typically using similar references. One major exception to this is  $b\to s\gamma$. Here all three  codes are in decent qualitative agreement (within a factor of $\sim 2$), but a more detailed comparison is difficult because the treatments of $b\to s\gamma$ is quite different in all three cases.  In \FF,  the latest, full NNLO results are used \cite{Misiak:2015xwa}.  \SF\ follows NLO results from `96 \cite{Chetyrkin:1996vx}, further enfolding hardcoded, unpublished SUSY loop calculations performed by the authors.  \FK\ follows the partial NNLO calculation of \cite{Misiak:2006ab,Lunghi:2006hc}.

In the remainder of this appendix, we will focus on the observables that show the largest discrepancies between the three codes. These are: $\Delta m_K$ and $\epsilon_K$, $\Delta m_{B_q}$, and $B_q\to\mu^+\mu^-$. 

\subsection{In depth comparison: $\Delta m_K$ and $\epsilon_K$}

 The sign of $\Delta m_K$ is physical relative to the sign of $\Delta \Gamma$ between $K_S$ and $K_L$.   \FF\ and \SF\  find a minus sign on the LLRR line, whereas \FK\ does not.  By examining the code, we find that \FK\ is  taking the absolute value of $\Delta m_K$. 

For $\epsilon_K$, \FF\ and \SF\ are in excellent agreement, while \FK\ disagrees substantially.  Part of the reason for this is that \FK\ is dividing by the theoretical prediction for $\Delta m_K$.  In most definitions of $\epsilon_K$, e.g.~\cite{Buras:2010pza}, it is divided by the very well known experimental value, a convention which both \FF\ and \SF\  follow.   This choice results in the possibility that $\epsilon_K$ values can, in principle, be larger than one, but obviously, this is a very sick region of parameter space with either definition. 

To facilitate an apples-to-apples comparison of \FF\ and \FK, we made the following changes to the latter: we modified the source code to divide $\epsilon_K$ by $(\Delta m_K)_{exp}$ instead of $(\Delta m_K)_{th}$, and we changed \FK's hadronic input parameters (the ``bag parameters" and $R_K$) to match \FF.  The result of this modification to \FK's result is shown in red dashed in figs.~\ref{fig:finalLLRR} and \ref{fig:finalLR}. As explained above, \FK\ is not performing any QCD RG between the SUSY scale and the low scale, so we should compare red dashed against blue dashed, which is \FF\ with QCD RG turned off.

While doing this brings the LLRR line into better agreement, the agreement with the LR line actually worsens. This implies there is a mismatch between \FF\ and \FK\ at the level of the Wilson coefficients themselves.   We have checked our analytic Wilson coefficients (gluino boxes) for against the MIA expressions of Gabbiani et al \cite{Gabbiani:1996hi} and found excellent agreement. This is shown in figs.~\ref{fig:finalLLRR} and \ref{fig:finalLR} in orange dashed. We also agree well with \SF,  so we conclude there is likely a mistake in the treatment in \FK. 
 
\subsection{In depth comparison: $\Delta m_{B_q}$}

For $\Delta m_{B_q}$, \FK\ disagrees with \SF\ and \FF\ along the LLRR line, but this disagreement is at most a factor of two.  However, along the LR line \FK\  disagrees immensely. 

 Most of this discrepancy is due to the two-loop double-Higgs penguins included in \FK.  To make a improved comparison, we have turned off these double-Higgs penguins (DHPs) in \FK's treatment of $\Delta m_{B_q}$ by modifying the source code (shown with the red dashed lines in figs.~\ref{fig:finalLLRR} and \ref{fig:finalLR}). We see that impact on the the LLRR line is small, but the agreement along the LR line becomes much better.    Thus, the DHPs in \FK\ are quite significant.  As $\tan\beta$ is not that large, and the heavy Higgs states are in the decoupling limit, this result is a bit surprising.  Checking against analytic expressions for the DHPs \cite{Buras:2001mb}, we find that the DHPs should be several orders of magnitude smaller than the values determined by \FK.
 
 Even with the DHPs  removed by hand from \FK,  the result has a noticeably different shape than that found in \FF\ and \SF, which agree well in their shapes.  There is more to the discrepancy than just what can be accounted for by the DHPs. 
 
 \subsection{In depth comparison: $B_q\to\mu^+\mu^-$}
 
 For $B_q\to\mu\mu$, \FK\ and \FF\ are in good agreement. \SF\ also agrees along the LLRR line, but differs qualitatively  along the LR line.   Along the LR line, both the SM and new physics contribution enter the branching ratio  dominantly in $F_A$, while for the LLRR line the SM contribution is in $F_A$, but the new physics contribution is mostly in $F_P$. From (\ref{eq:BRbqmumu}), we can see that,
\begin{equation}
\text{BR}(B_i\rightarrow \mu^+ \mu^-) \propto \left| {F}_{A,SM}+ {F}_{A,NP} + {F}_{P,NP} \right|^2.
\end{equation}
Thus there is a relative sign discrepancy between the two codes in ${F}_{A,SM}$ and  $F_{A,NP}$ in the LR case.  Using \FF, we can compute both the new physics and SM piece in situ to conclusively determine the relative sign between the two, since all other factors are treated identically between the two pieces.  For this reason, we are confident in the sign found with \FF. (And \FK\ confirms it.) In \SF\, both the SM and new physics contributions are hard-coded formulas, so a mistake in the overall sign could very plausibly have been introduced.

\bibliography{FFmanual}

\printindex

\end{document}